\documentclass[a4paper,noarxiv, twocolumn, accepted=2024-08-20]{quantumarticle}
\usepackage[utf8]{inputenc}
\usepackage[english]{babel}
\usepackage[T1]{fontenc}

\usepackage{empheq}
\usepackage{float}
\usepackage{lipsum}
\usepackage{mathtools, cuted}
\usepackage{esvect}
\usepackage{xcolor}
\usepackage[caption = false]{subfig}
\usepackage{graphicx}
\usepackage{dcolumn}
\usepackage{bm}
\usepackage{hyperref}
\usepackage{physics}
\usepackage[mathlines]{lineno}
\usepackage[capitalise]{cleveref}
\usepackage[english]{babel}
\usepackage{orcidlink}

\usepackage[numbers]{natbib}

\newcommand{\bk}[1]{\langle #1 \rangle}

\begin{document}

\title{The Fractal-Lattice Hubbard Model}

\author{Monica Conte\,\orcidlink{0009-0000-2632-6306}}
\affiliation{Institute for Theoretical Physics, Utrecht University,
Princetonplein 5, 3584CC Utrecht, The Netherlands,}
\author{Vinicius Zampronio\,\orcidlink{0000-0001-8236-9052}}
\affiliation{Departamento de Física Teórica e Experimental - UFRN, Av. Sen. Salgado Filho 3000, 59078-970, Natal - RN, Brazil,}
\affiliation{Institute for Theoretical Physics, Utrecht University,
Princetonplein 5, 3584CC Utrecht, The Netherlands,}
\author{Malte Röntgen\,\orcidlink{0000-0001-7784-8104}}
\affiliation{Laboratoire d’Acoustique de l’Université du Mans,
Unite Mixte de Recherche 6613, Centre National de la Recherche Scientifique,
Avenue O. Messiaen, F-72085 Le Mans Cedex 9, France}
\author{Cristiane Morais Smith\,\orcidlink{0000-0002-4190-3893}}
\affiliation{Institute for Theoretical Physics, Utrecht University, Princetonplein 5, 3584CC Utrecht, The Netherlands,}

\date{\today}

\begin{abstract}
	Here, we investigate the fractal-lattice Hubbard model using various numerical methods: exact diagonalization, the self-consistent diagonalization of a (mean-field) Hartree-Fock Hamiltonian and state-of-the-art Auxiliary-Field Quantum Monte Carlo. We focus on the Sierpinski triangle with Hausdorff dimension $1.58$ and consider several generations. In the tight-binding limit, we find compact localised states, which are also explained in terms of symmetry and linked to the formation of a ferrimagnetic phase at weak interaction. Simulations at half-filling revealed the persistence of this type of magnetic order for every value of interaction strength and a Mott transition for U/t $\sim$ 4.5. In addition, we found a remarkable dependence on the Hausdorff dimension regarding $i)$ the number of compact localised states in different generations, $ii)$ the scaling of the total many-body ground-state energy in the tight-binding limit, and $iii)$ the density of the states at the corners of the lattice for specific values of electronic filling. Moreover, in the presence of an intrinsic spin-orbit coupling, the zero-energy compact localized states become entangled and give rise to inner and outer corner modes.
\end{abstract}
\maketitle

\renewcommand{\figurename}{FIG.}

\section{\label{sec:Intro}Introduction}

The one-band Hubbard model \cite{Intro_Hubbard} is one of the simplest and most fundamental models for describing the effect of electron correlation in solids. In its fermionic version, it physically resembles electrons hopping in a lattice configuration, with hopping parameter $t$ and on-site interaction parameter $U$.
It has been solved analytically in one-dimension \cite{Lieb} using an extension of the Bethe ansatz technique \cite{Bethe1,Bethe2}. However, despite its deceptively simple form, analytical solutions in higher dimensions are yet to be found. In this case, quantum simulations~\cite{Chiu2018,Tusi2022,Yang2022}, machine learning~\cite{Chiu2019}, and numerical techniques such as Density Matrix Renormalization Group (DMRG) \cite{spinliq,dmrg2,sza21,Zhu2022}, Linked cluster expansion~\cite{gar22}, Variational Monte Carlo \cite{Becca, Becca2,toc21}, Quantum Monte Carlo (QMC) \cite{qmc1, Vinicius}, among others~\cite{wie21} must be implemented.

The relevance of the model and its versatility are demonstrated by its numerical applicability to a wide range of two-dimensional lattice structures, leading to the exploration of numerous phases of matter. For instance, the Hubbard model is able to describe spin and density waves \cite{waves} and, in frustrated geometries, spin liquids \cite{shi17,spinliq,Becca2,sza21,toc21,che22}. In addition, the model is related to unconventional superconductivity \cite{hightsuper2,Zhu2022,Vinicius}. 

Despite the progress made in more than $60$ years of its discovery, the study of the Hubbard Model in a non-integer dimension still remains an open problem. Fractals offer a way to explore lattice configurations with non-integer dimensions. These geometrical structures are characterized by their intricate and self-replicating patterns, obtained through iterative processes, in which a basic shape is repeatedly transformed or replicated. 

Recently, fractal geometries have been successfully realized in experimental settings, offering empirical validation of theoretical concepts. A fractal lattice with electrons has been engineered and characterized \cite{Kempkes}. Quantum transport in fractal photonic lattices \cite{quantum_transport} revealed anomalous behaviour, deviating from the expected patterns observed in infinite regular lattices. In addition, higher-order topological insulators have been theoretically \cite{Floq_theor1,Floq_theor2} and experimentally \cite{Floq_exp} observed in acoustic experiments on a Sierpinski carpet, in which the Hausdorff dimension is $d_{H}\sim 1.89$. Finally, topological corner modes were shown to emerge not only in the above mentioned metamaterials, but also in real materials such as thin layers of bismuth deposited on InSb substrates \cite{Robert}.

Here, we investigate the Hubbard model in a fractal lattice. We focus on the Sierpinski triangle, which has a Hausdorff dimension of $d_{H} = \log(3)/\log(2) \sim 1.58$.
This lattice geometry is non periodic, which challenges conventional approaches, such as band theory, and requires the exploration of alternative methods to analyze the system. Moreover, the fractal lattice that we adopt contains sites with different connectivity and is bipartite, uniquely influencing the particle behaviour. 
We first study the model in the limit of zero interaction, since we expect the geometry of the lattice to influence mainly the kinetic term. Afterwards, we introduce interaction and employ three distinct numerical methods to find the ground-state solution: exact diagonalization, mean-field Hartree-Fock approximation, and Constrained-Path Auxiliary-Field Quantum Monte Carlo (CP-AFQMC). Among these methods, we predominantly implement CP-AFQMC due to the exact nature of QMC methods. For fermionic ground-states, a CP approximation is often required to deal with the infamous sign problem \cite{sign, sign2}. 

The study of the system without interaction allowed to understand the consequence of the fractal geometry at a tight-binding (TB) level. In particular, we find the presence of so-called compact localized states (CLS), that is, states which strictly vanish on a large number of sites due to destructive interference \cite{Leykam2018AP31473052ArtificialFlatBandSystems}.
We notice that the number of these states scales as the Hausdorff dimension when the generation of the fractal lattice increases. This scaling property is also shared by the total many-body ground-state energy. Moreover, we observe that the average density at the corners of the lattice corresponds to $d_{H}$ for a specific value of electronic filling. In the presence of intrinsic spin-orbit coupling, the degenerate CLS at zero energy become entangled and lead to the formation of robust corner modes. From the implementation of the Hubbard model on the lattice we identify both a Mott transition and a ferrimagnetic phase. The latter, for weak interaction, can be linked to the observations made in the TB limit. 

This paper is structured as follows: in \cref{sec:Model}, we introduce both the fractal lattice and the Hubbard Hamiltonian; we list their properties relevant to this work and the methods that we used to solve the problem. \cref{sec:TBresults} is dedicated to the results found with the TB approach at $U=0$. Understanding these results requires a symmetry-based analysis, explained in \cref{sec:CLS}, which reveals interesting results related to the fractality of the lattice. In \cref{sec:ISOC}, we investigate the effect of an intrinsic spin-orbit coupling on the degenerate CLS. Finally, the study of the quantum phases of the system obtained with CP-AFQMC is presented in \cref{sec:QMC}. Our conclusions are presented in \cref{sec:Concl}.

\section{\label{sec:Model}The Model}
\begin{figure}[t]
	\centering
	\includegraphics[width=\columnwidth]{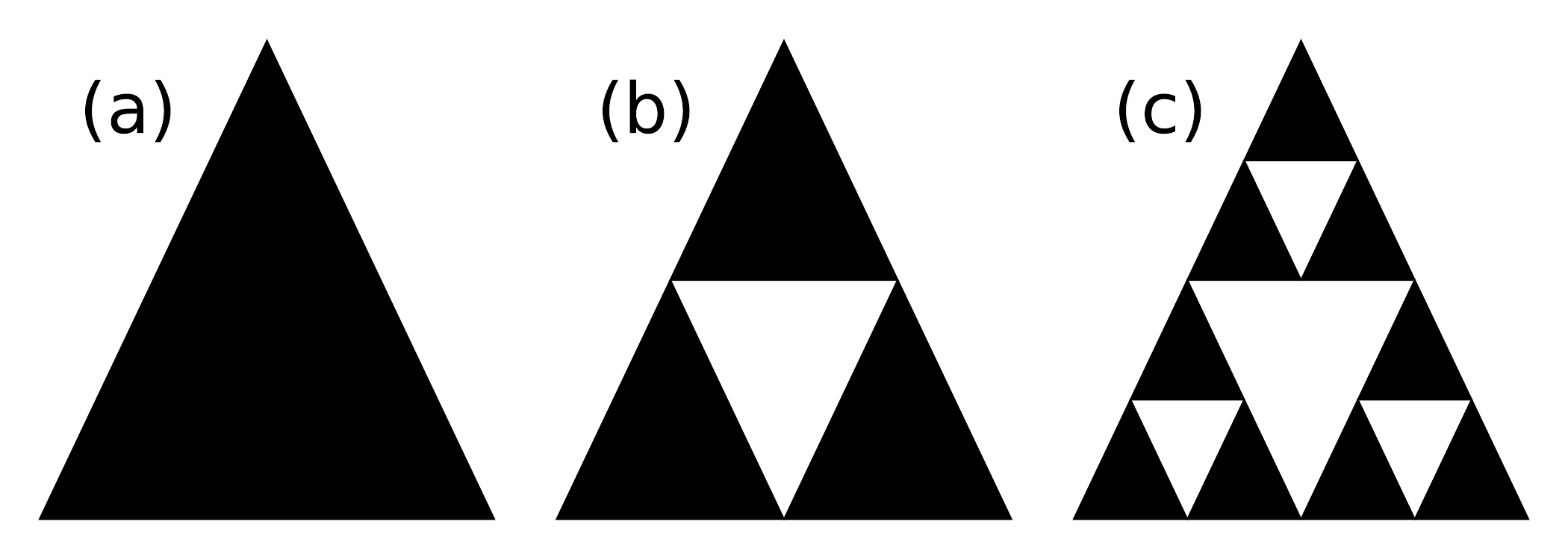}
	\caption{First three generations of the Sierpinski triangle. (a) Generation zero, (b) first generation, and (c) second generation.}
	\label{fig:Striangle}
\end{figure}

\begin{figure}[b]
	\centering
	\includegraphics[width=0.9\columnwidth]{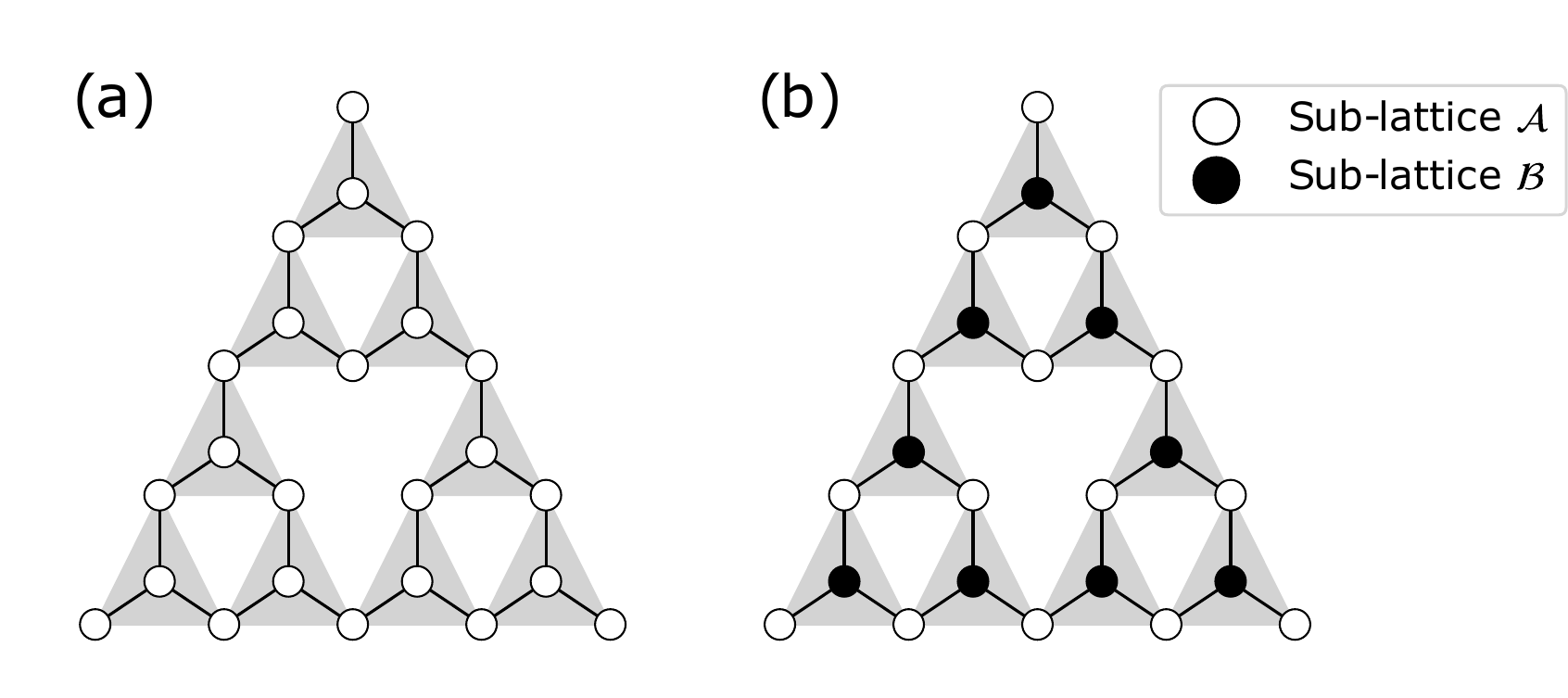}
	\caption{(a) Second generation of the fractal lattice, (b) division of the lattice into two sublattices.}
	\label{fig:fractal_lattice}
\end{figure}

Let us start by constructing the fractal lattice from the Sierpinski triangle. \cref{fig:Striangle} shows the first three generations of this geometrical structure.
Here, we focus on the second generation. One possible way to construct the fractal lattice is by placing lattice sites in the centre and in the corners of the remaining triangles, as shown in \cref{fig:fractal_lattice}. Moreover, these sites are linked in such a way that all the links between sites have the same length.
We colour the sites according to their connectivity. Sites at the bulk of the triangles are coloured black [see \cref{fig:fractal_lattice}(b)] and the boundary sites are white. Thus, black sites have connectivity $3$, while white sites have connectivity $2$, except for the sites at the corners of the triangle, which have only one neighbour.


We study the Hubbard model with Hamiltonian
\begin{equation}
	H = -t \sum_{\langle ij\rangle \in \Lambda} \sum_{\sigma=\uparrow, \downarrow} \left( c^{\dagger}_{i,\sigma} c_{j,\sigma} + \text{h.c.} \right) + U\sum_{i} n_{i,\uparrow}n_{i,\downarrow}.
	\label{eq:hubb_ham}
\end{equation}
Here, $i, j$ are site indices and $\sigma$ is a spin index, which refers to the up or down projection of the electron's spin along a fixed axis. The operator $n_{i,\sigma} = c^{\dagger}_{i,\sigma} c_{i,\sigma}$ is the number operator and one of the sums runs over nearest-neighbouring sites $\langle ij\rangle$ belonging to the set $\Lambda$ containing the sites indices. The first term on the RHS describes hopping of electrons on neighbouring sites, with hopping amplitude $t$. For simplicity, we set $t=1$ as the energy unit. The second term accounts for Coulomb interactions between electrons with opposite spin on the same lattice site, described by the Hubbard parameter $U$. We set open boundary conditions. 

Due to the bipartiteness of the lattice, \cref{fig:fractal_lattice}(b), the model is particle-hole symmetric at half-filling \cite{Intro_Hubbard}. In addition, the Hamiltonian has a global $SU(2)$ invariance \cite{Intro_Hubbard}, which reflects the spin-rotational symmetry. 

We are interested in the ground-state properties of the system. However, a naive exact diagonalization of the many-body Hamiltonian would require the diagonalization of a matrix of size $4^{M} \times 4^{M}$, with $M$ the number of sites. This is computationally demanding, and we could not go beyond the first generation. In the following sections, we apply various methods to overcome this difficulty. \\

\section{\label{sec:TB}Tight-Binding Approach}
In order to study the many-body ground-state properties, we follow an approach that, in the framework of second quantization, makes use of the TB limit.
Therefore, we start by investigating the Hubbard model in the limit when the interaction parameter $U=0$. The resulting Hamiltonian represents electrons hopping freely around the lattice, without any energy cost for double occupation of lattice sites. 

Given $N$ particles that populate the system, one builds the possible single-particle orbitals as superpositions of single-site wavefunctions. These orbitals, in turn, are used as basis of the $N$-particle Fock space, where we are able to construct anti-symmetrized many-body wavefunctions making use of Slater determinants. We follow closely the formalism detailed in Ref. \cite{Nguyen}. 

We also study the Hamiltonian at the single-particle level, since the many-body behaviour is strictly related. This means that we perform a spectral analysis of the single-particle orbitals that we use to construct the Slater determinant. Notice that computing these orbitals simply consists in diagonalizing the Hamiltonian $H_{TB}$ in the single-particle basis, on which it reduces to the size $M\times M$.

\subsection{\label{sec:TBresults}Tight-binding ground-state properties}
\begin{figure}[t]
	\centering
	\includegraphics[width=0.9\columnwidth]{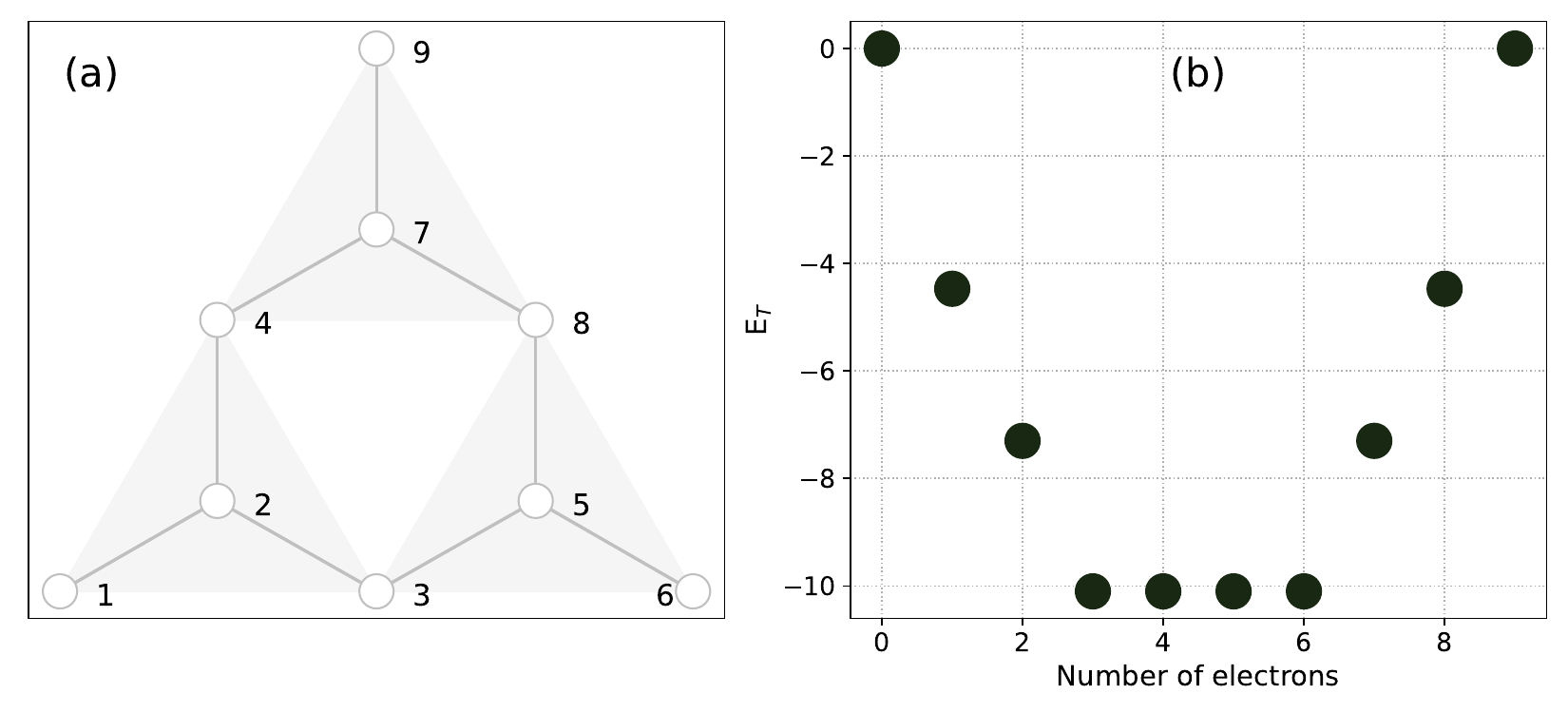}
	\caption{(a) First generation of the fractal lattice. (b) Its corresponding many-body ground-state energy for different values of $N_{\sigma}$. The $x$ axis in (b) indicates the number of electrons of one kind $\sigma$. The total number of electrons in the system is $2N_{\sigma}$.}
	\label{fig:lattice_energytot1gen}
\end{figure}

In this section, we present results of TB implementations on the fractal lattice. In \cref{fig:lattice_energytot1gen}, we show the first generation of the lattice and the behaviour of the ground-state many-body energy as a function of electronic filling $N = 2 N_{\sigma}$, where $N_{\sigma}=N_{\uparrow}=N_{\downarrow}$ is the number of electrons with spin $\sigma = \uparrow,\downarrow$.

\begin{figure}[t]
	\centering
	\includegraphics[width=0.9\columnwidth]{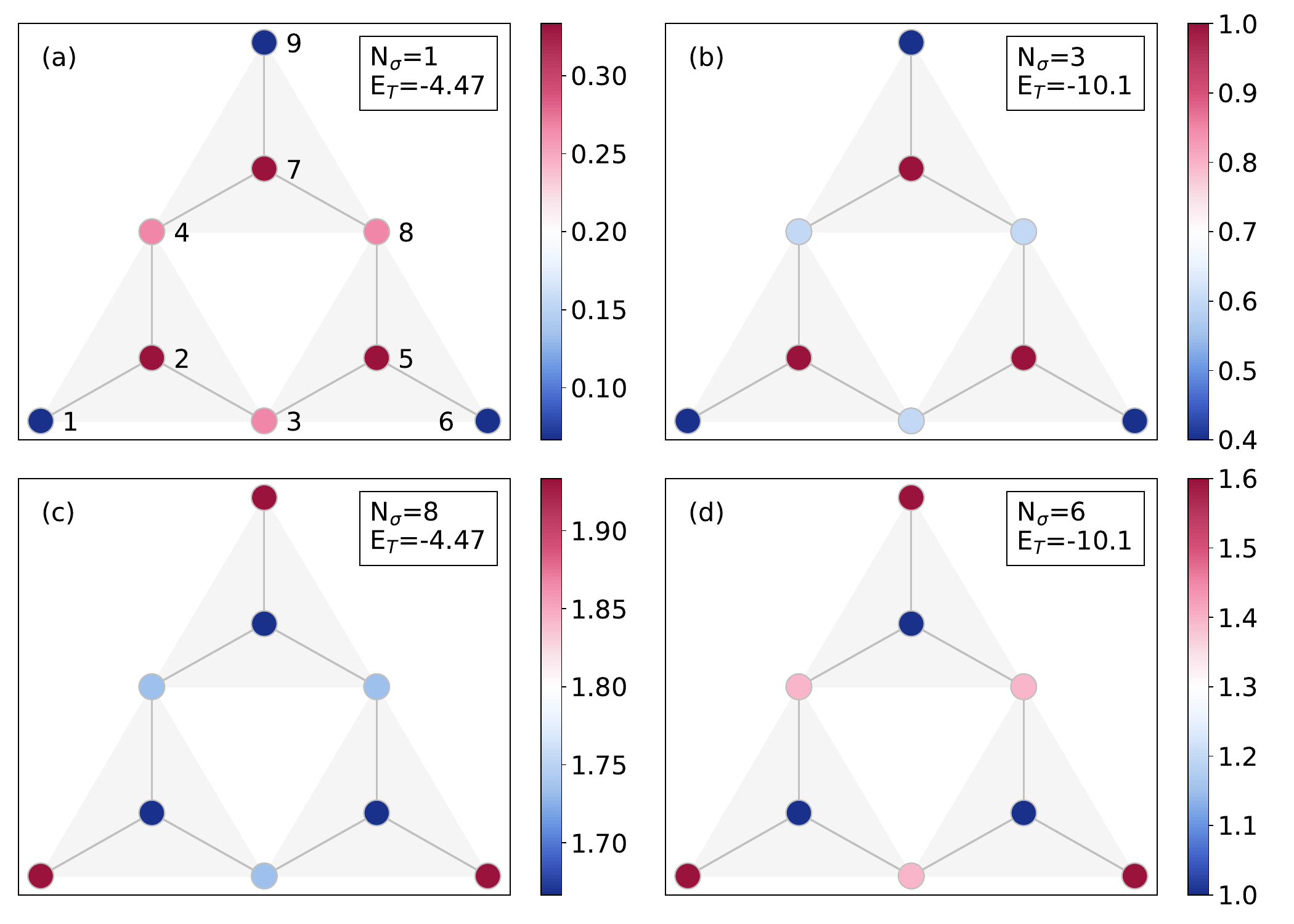}
	\caption{Average density configuration of the TB ground-state on the first generation of the fractal lattice with filling (a) $N_{\sigma}=1$, (b) $N_{\sigma}=3$, (c) $N_{\sigma}=8$ and (d) $N_{\sigma}=6$. Notice the different scale in the colour bar for each individual picture.}
	\label{fig:quattro1gen}
\end{figure}

The symmetry in the energy distribution reflects the fact that at zero interaction, the system is particle-hole symmetric around half-filling. This means that the system is equivalent upon exchange of particles with holes and vice versa. This symmetry is also evident in the configurations displaying the average density per site on the lattice
\[
\bk{n_{i}} = \bk{n_{i,\uparrow}} + \bk{n_{i,\downarrow}}
\]
shown in \cref{fig:quattro1gen}. Comparing the configurations with $N_{\sigma}=1$ [\cref{fig:quattro1gen}(a)] and $N_{\sigma}=8$ [\cref{fig:quattro1gen}(c)], we conclude that electrons are placed in an empty lattice in the same way that holes are placed in a fully-filled lattice. Therefore, configuration with $N_{\sigma}=8$ and $N_{\sigma}=1$ are the inverse of each other since there are $9$ sites and $N=8=9-1$.
Notice that, below half-filling, the sites with higher connectivity, i.e. the three sites with three neighbours at the centre of the triangles, have a higher density. We can understand this behaviour by taking into account the fact that hopping is energetically favourable, so in the ground-state it is preferable to store electrons in sites from which they have more probability to hop around. Along this logic, the sites with connectivity $2$ are less populated, followed by the sites at the corners, which are connected to the lattice just through one link. If we increase the number of electrons and place $N_{\sigma}=3$ electrons as in \cref{fig:quattro1gen}(b), the Pauli exclusion principle starts to play a role, meaning that two electrons with same spin cannot occupy the same lattice site. The electrons then must occupy sites which are kinetically less favourable, if the most favourable ones are already occupied.
When we exceed half-filling, i.e. the case where the total number of electrons coincides with the number of sites, the pattern is inverted: the three corners are then more populated, followed by the sites with connectivity $2$ and finally the sites with connectivity $3$, as shown in \cref{fig:quattro1gen}(d). It seems energetically more favourable to store electrons in the corners and let the ones in the bulk hop. Configurations above half filling follow this pattern, see for example \cref{fig:quattro1gen}(c). 

\subsection{\label{sec:CLS}Compact Localized States and Scaling Properties}
In order to understand more in detail the behaviour of the system, we must determine the eigenvalues $E_{n}$ and eigenvectors $\chi_{n}$ of the single-particle Hamiltonian. In the TB limit, spin-up and spin-down sectors are decoupled, and we can thus treat them independently. 

\subsubsection{First generation}
The first generation of the fractal lattice has $9$ sites and the Hamiltonian is therefore a $9\times 9 $ matrix.
\begin{figure}[t]
	\centering
	\includegraphics[width=0.73\columnwidth]{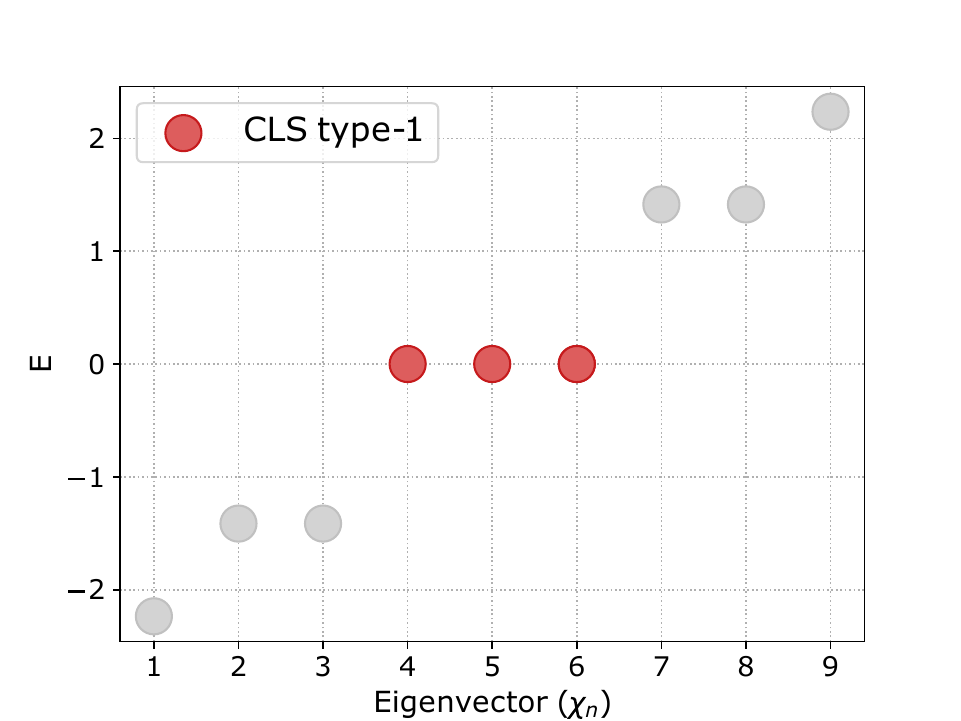}
	\caption{Spectrum of the single-particle TB Hamiltonian for the first generation of the fractal lattice.
		The spectrum shows different energy levels: one three-fold degenerate level at zero energy, two two-fold degenerate levels at $E_{n}=\pm \sqrt{2} $ and two non-degenerate levels.}
	\label{fig:energyspec1gen}
\end{figure}
The energy spectrum is symmetric with respect to $E=0$, see \cref{fig:energyspec1gen}.
This is a direct consequence of the bipartiteness (chiral symmetry) of the system.
This bipartite character is maintained for higher generations of the fractal, and consequently their energy spectra are symmetric around $E = 0$ as well.

\begin{figure}[b]
	\centering
	\includegraphics[width=\columnwidth]{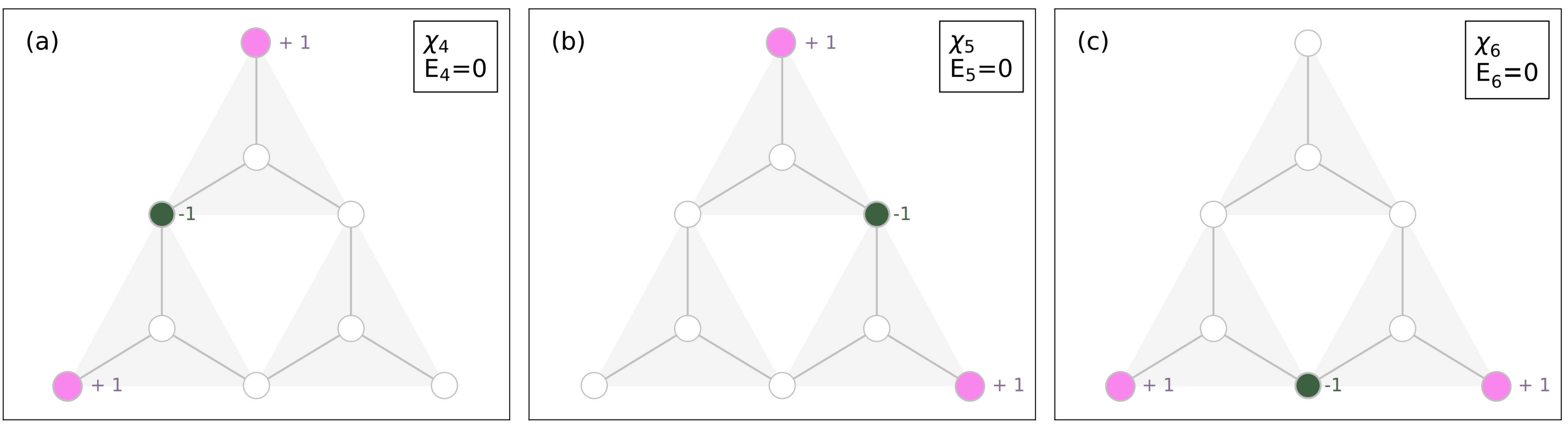}
	\caption{Amplitude scaled to unity of the three basis eigenvectors in the zero-energy level of the TB Hamiltonian, in (a) $\chi_4$, in (b) $\chi_5$ and in (c) $\chi_6$. Dark-green and pink dots on the lattice represent, respectively, negative and positive amplitude of the wavefunction on the sites. On sites with white dots, the amplitude is zero.}
	\label{fig:3basis}
\end{figure}

Let us now consider the eigenvectors $\chi_4, \chi_5$, and $\chi_6$ belonging to the three-fold degenerate level $E=0$. \Cref{fig:3basis} shows their amplitude on the sites of the lattice. 
An interesting behaviour becomes evident: the amplitudes on sites that belong to the sublattice $\mathcal{B}$ are zero. This is a consequence of the fact that the amplitudes of their neighbouring sites sum up to zero and lead to destructive interference \cite{Leykam2018AP31473052ArtificialFlatBandSystems}. 
These states are examples of so-called CLS, since they are perfectly localized on a specific set of sites in the lattice, such that their amplitude on the rest of the lattice is exactly zero.

Before continuing, let us remark two interesting points about such states.
Firstly, CLS are completely robust against any perturbations---no matter how strong---that only affect the sites on which they vanish \cite{Rontgen2019PRL123080504QuantumNetworkTransferStorage}.
This makes these states a candidate for the storage of information in the form of qubits \cite{Kempkes2023QF21CompactLocalizedBoundaryStates,Rontgen2019PRL123080504QuantumNetworkTransferStorage}.
Secondly, in periodic systems, CLS usually lead to the emergence of one or more perfectly flat bands, which recently became a topic of intense research interest; see Refs. \cite{Leykam2018AP31473052ArtificialFlatBandSystems} and \cite{Leykam2018AP3070901PerspectivePhotonicFlatbands} for two reviews on flat-band systems.

The existence of the three zero-energy CLS, $\chi_4, \chi_5$ and $\chi_6$, can also be deduced as follows: For a bipartite system with $N_\mathcal{A}$ ($N_\mathcal{B}$) sites in the $\mathcal{A}$ ($\mathcal{B}$) sublattice, it is a classical result that there must be at least $\Delta N = |N_\mathcal{A} - N_\mathcal{B}|$ zero-energy eigenstates which have vanishing amplitudes on the minority sublattice \cite{Sutherland1986PRB345208LocalizationElectronicWaveFunctions,Lieb1989PRL621201TwoTheoremsHubbardModel}. In our case, we have $N_\mathcal{A} = 6$, $N_\mathcal{B} = 3$, yielding $\Delta N = 3$, which is exactly the number of zero-energy CLS that we found.

\begin{figure}[t]
	\centering
	\includegraphics[width=0.73\columnwidth]{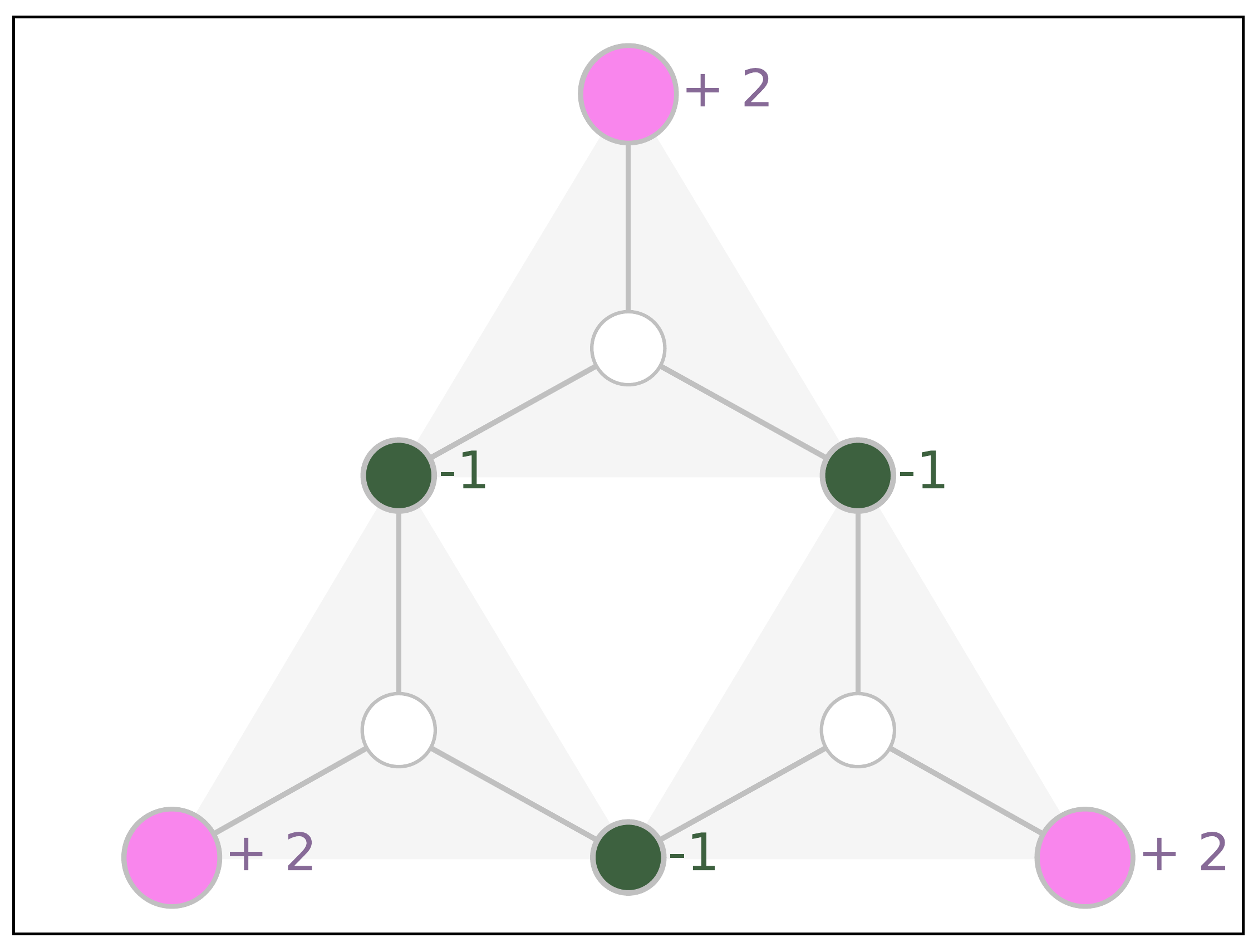}
	\caption{Wavefunction that exemplifies how states around half-filling can have a larger contribution of the average density to the corners. It is obtained by combining the states belonging to the basis of the zero-energy degenerate level $\chi_4 + \chi_5 + \chi_6$.}
	\label{fig:sumcorners}
\end{figure}

We are now able to better understand the behaviour of zero-energy states: the three CLS $\chi_4, \chi_5$ and $\chi_6$ discussed above form a basis in the degenerate subspace, meaning that a general eigenstate with eigenvalue zero is found by taking linear combinations of $\chi_4, \chi_5$, and $\chi_6$. Any pair of these three states share one corner.
For instance, eigenvectors $\chi_4$ and $\chi_5$ share the top corner; a linear superposition $\chi_4 + \chi_5$ will thus have enhanced amplitude at this shared corner.
Equipped with these insights, we can therefore understand why states around half-filling contribute to the density in the corners. As an example to make this point clearer, we show the density configuration $\chi_4 + \chi_5 + \chi_6$ in \cref{fig:sumcorners}.

Importantly, the existence of CLS and the subsequent density-enhancement at the corners of the Sierpinsky triangle is not limited to a simple TB picture.
Indeed, as we show in \cref{sec:ISOC}, it persists even when considering intrinsic spin-orbit coupling. Moreover, the scaling of the number of CLS---as investigated further below---shows the same behaviour in both cases, that is, with and without intrinsic spin-orbit coupling.

Having dealt with the CLS, let us now turn our attention to the other eigenstates, in particular, to the two-fold degenerate ones. 
Their degeneracy is a consequence of the $C_{3v}$ point-symmetry group \cite{Cotton1990ChemicalApplicationsGroupTheory}, which has order $6$ and contains the identity, two rotations $C_{3}, C_{3}^2$ and three reflections. This group has an irreducible representation of dimension $2$, which explains two-fold degeneracies \cite{Landau}.
As a side note, these levels do not posses the property of CLS states. There is not a definite group of lattice sites where the amplitude is zero for every eigenvector of the degenerate level, as it was the case for the zero-energy states.

It is now possible to look at the density configurations in the first generation of the fractal lattice, \cref{fig:quattro1gen}, from a different perspective. Configurations below the degenerate level around half-filling populate the sites belonging to sub-lattice $\mathcal{B}$ with an average density $\bk{n_i}=1$, see \cref{fig:quattro1gen} (b). This is in agreement with the fact that, due to particle-hole symmetry, the average density per site at half-filling is $\bk{n_i}=1$ and the zero-energy CLS only populate sublattice $\mathcal{A}$. Starting from the configuration with $N_{\sigma}=3$ and raising the number of electrons until $N_{\sigma}=6$ means including zero-energy CLS in the many-body wavefunction. 
The total many-body energy does not increase because the CLS added have zero energy, but sites belonging to sublattice $\mathcal{A}$, in particular the corners, become more populated.

In the following, we analyze the emergence of CLS in higher generations of the fractal.
We shall see that there appear more CLS, both at zero and finite energy.

\subsubsection{Second generation}
\begin{figure}[b]
	\centering
	\includegraphics[width=0.74\columnwidth]{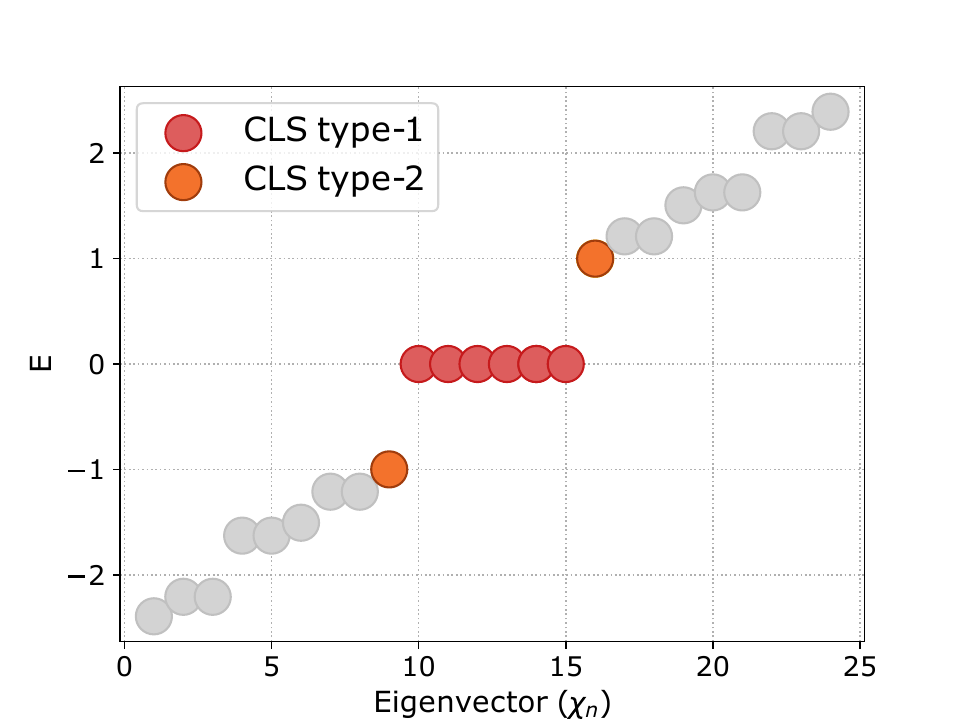}
	\caption{Spectrum of the single-particle TB Hamiltonian for the second generation of the fractal lattice. The spectrum shows different energy levels: a six-fold degenerate level (CLS type-1) at zero energy, two states (CLS type-2) at energy $\pm 1$ and non-CLS eigenstates, the latter represented by grey dots.}
	\label{fig:energyspec2gen}
\end{figure}

The energy spectrum of the Hubbard Hamiltonian on the second-generation of the fractal lattice is shown in \cref{fig:energyspec2gen}.
Once again, due the fractal's bipartiteness, the energy spectrum is symmetric with respect to $E = 0$.
Now, there are $N_\mathcal{A}= 15$ and $N_\mathcal{B} = 9$ sites in the two respective sublattices, which results in $|N_\mathcal{A} - N_\mathcal{B}| = 6$ eigenstates with zero energy.

\begin{figure}[t]
	\centering
	\includegraphics[width=0.9\columnwidth]{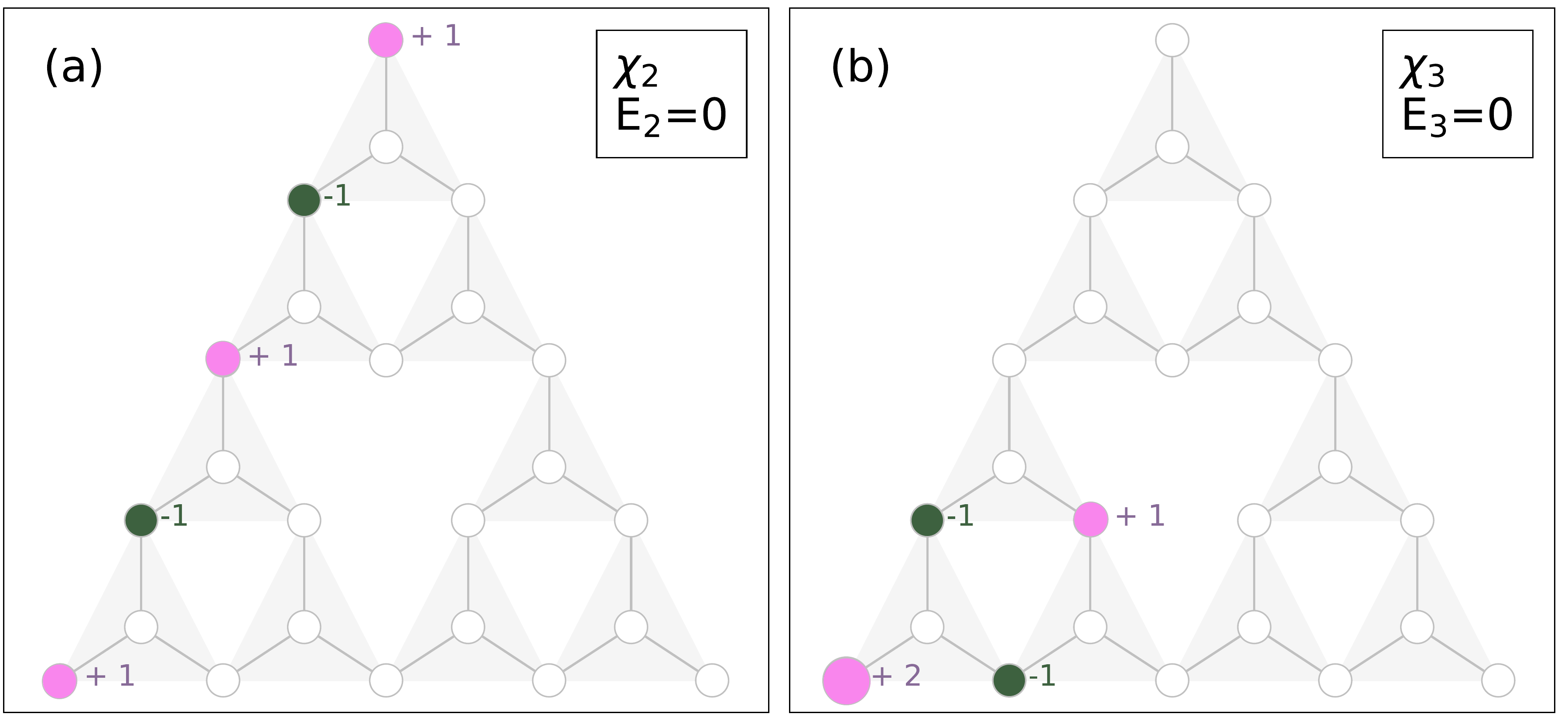}
	\caption{Prototype basis states in the zero-energy degenerate space of the second generation of the fractal lattice.}
	\label{fig:basis2gen}
\end{figure}

Interestingly, these six zero-energy CLS can be generated from just two prototypical states, show in \cref{fig:basis2gen}.
To obtain six states from these two, each of them must be rotated by $120^{\circ}$ and $240^{\circ}$.
The resulting set of six states are linearly independent, and thus span the entire six-dimensional degenerate subspace of eigenvalue zero.

\begin{figure}[b]
	\centering
	\includegraphics[width=0.9\columnwidth]{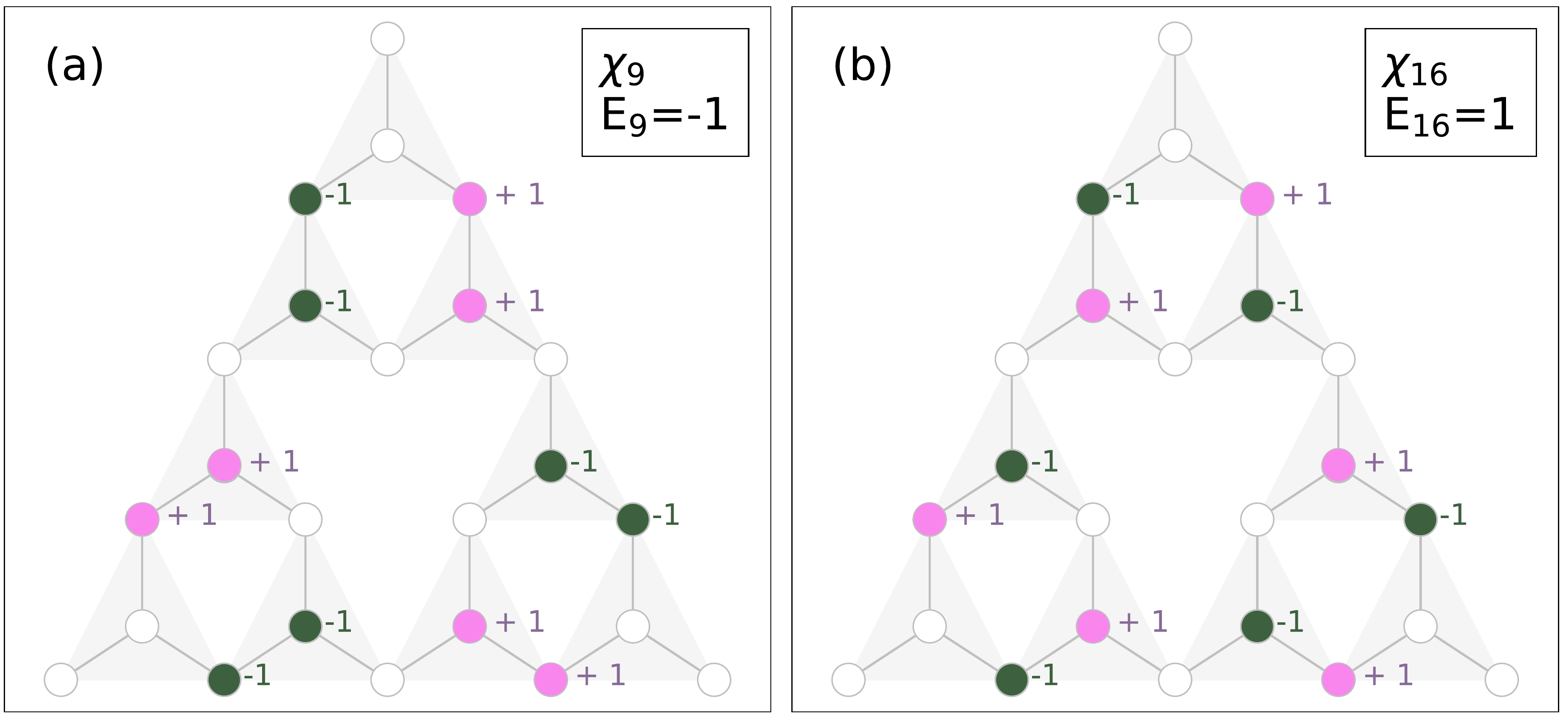}
	\caption{Eigenstates corresponding to the eigenvalues of the two-fold unitary degenerate level CLS type-2. (a) One of the eigenvectors with energy $E=-1$. (b) One of the eigenvectors with energy $E=1$. They are the second type of CLS found by spectral analysis.}
	\label{fig:oneenergy2gen}
\end{figure}

Apart from these six zero-energy states, the second generation now features a new type of CLS at the non-degenerate energy levels $\pm 1$.
These states are depicted in \cref{fig:oneenergy2gen}.
We note that these two states are tightly related to each other: One can be constructed from the other, simply by flipping the sign on all sites of the sublattice $\mathcal{B}$.
This is again a direct consequence of the fractal's bipartiteness.

As we shall see, in the third generation, CLS appear at even more energies than just $0$ and $\pm 1$.
To ease the discussion, we will enumerate the CLS according to the absolute value of their energies.
We denote the zero-energy CLS as type-1, and the ones at $E=\pm 1$ as type-2.

\subsubsection{Third generation}
\begin{figure}[t]
	\centering
	\includegraphics[width=0.8\columnwidth]{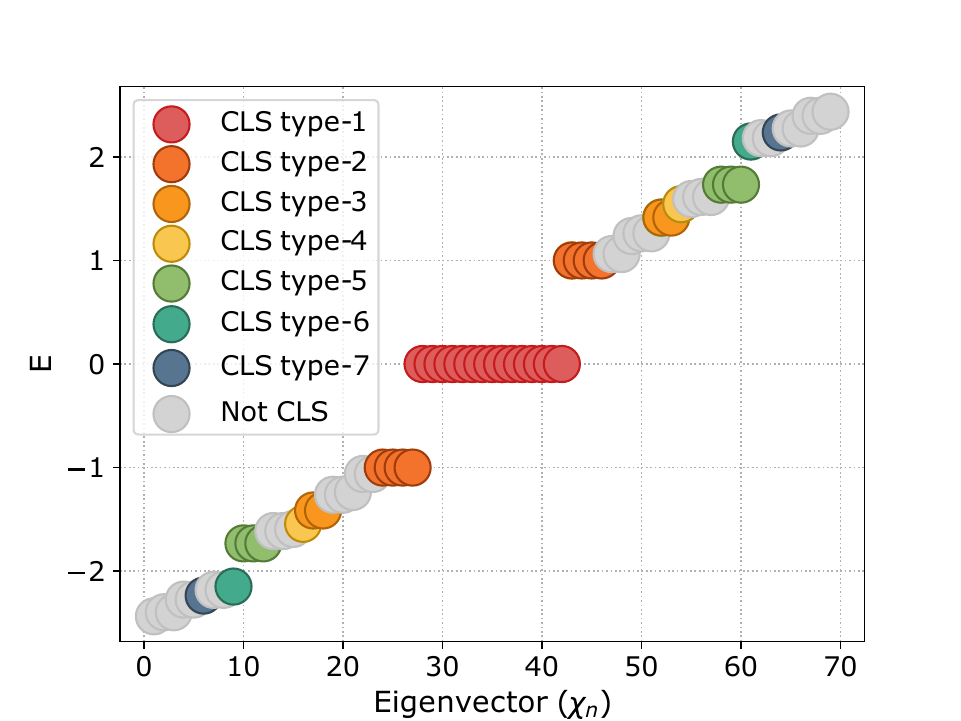}
	\caption{Spectrum of the single-particle TB Hamiltonian for the third generation of the fractal lattice. The spectrum shows seven different types of CLS: a $15$-fold degenerate type-$1$ CLS at zero energy, a $4$-fold degenerate type-$2$ CLS at $E=\pm 1$, a $2$-fold degenerate type-$3$ CLS at $E=\pm \sqrt{2}$, a type-$4$ CLS at $E=\pm 1.543$, a $3$-fold degenerate type-$5$ CLS at $E=\pm \sqrt{3}$, a type-$6$ CLS at energy $E=\pm 2.149$ and a type-$7$ CLS with energy $E=\pm \sqrt{5}$.}
	\label{fig:energyspec3gen}
\end{figure}

\begin{figure}[t]
	\centering
	\includegraphics[width=0.7\columnwidth]{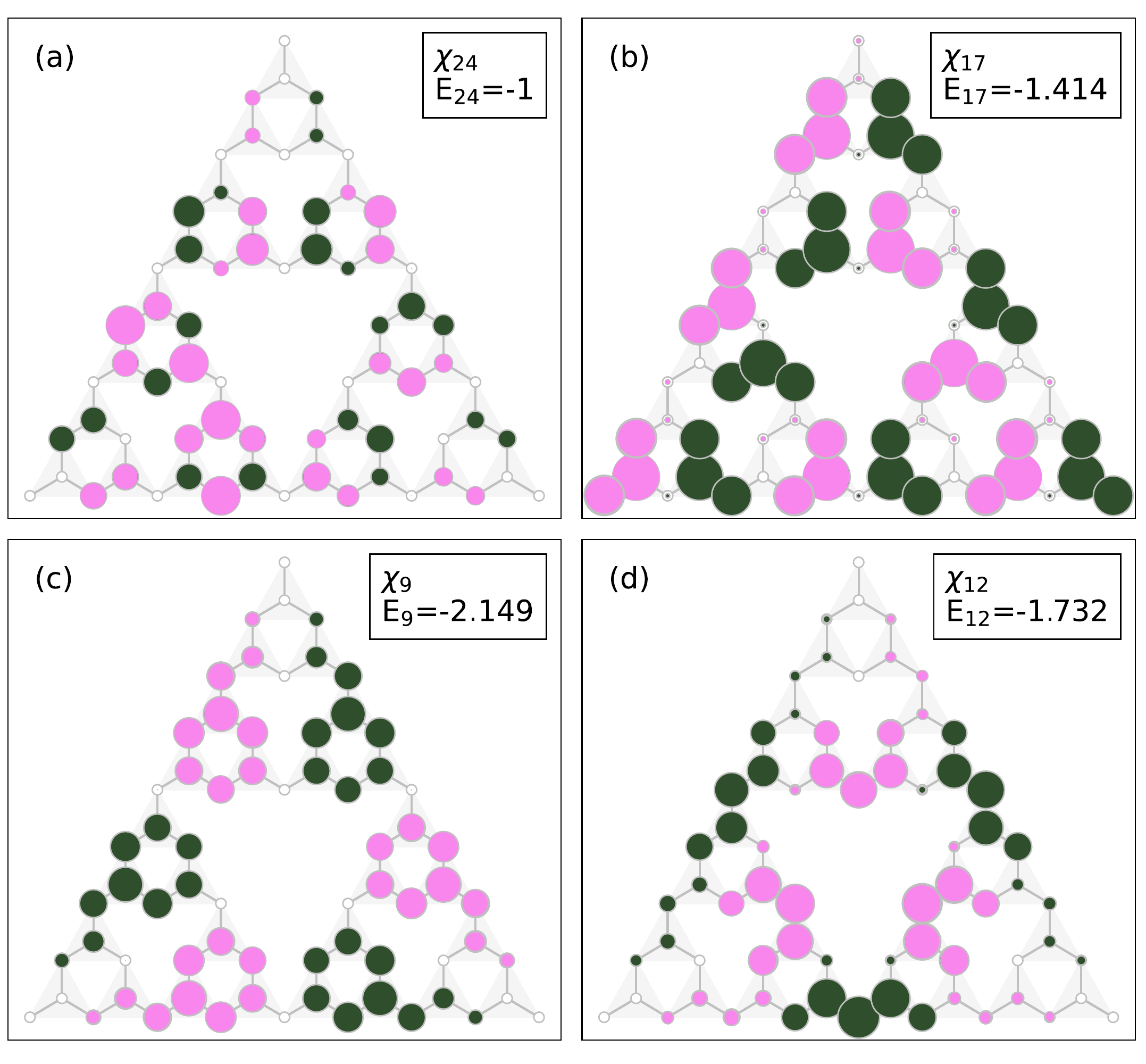}
	\caption{Four out of the seven types of CLS on the third generation of the fractal lattice. (a) One of the type-2 CLS with energy $E=-1$, 
		(b) one of the type-3 CLS with energy $E=-\sqrt{2}$ (c) one of the type-5 CLS with energy $E=-\sqrt{3}$, (d) one of the type-6 CLS with energy $E=-2.149$. The dark-green (pink) dots indicate sites with negative (positive) parity, where the size of the dots represents the value of the amplitude: the larger the dot, the higher the amplitude.
	}
	\label{fig:quattro3gen}
\end{figure}

The energy spectrum for the third generation TB model is shown in \cref{fig:energyspec3gen}.
Apart from the type-1 and type-2 CLS that we encountered in the second generation, there are now five additional types of CLS, namely, type-3 to type-7.
Examples of some CLS are depicted in \cref{fig:quattro3gen}.
Again, the fractal's bipartiteness means that each of the depicted states has a partner whose amplitudes on each site of the $\mathcal{B}$-sublattice is flipped, and which has the same energy, but with an inverted sign.

Let us now discuss some aspects of the CLS in more detail.
The (non-depicted) type-1 CLS are simple: they vanish on the entire sublattice $\mathcal{B}$.
We note that there are $N_{\mathcal{A}} = 42$ and $N_{\mathcal{B}}=27$ sites in the two sublattices, resulting in $42-27 = 15$ type-1 (zero-energy) CLS.
The CLS of type-2, 5, and 6 [\cref{fig:quattro3gen}(a), (c) and (d) respectively] have a very peculiar property: they vanish on all three other corners of the fractal.
On the other hand, the CLS of type-3 [depicted in \cref{fig:quattro3gen}(b)] and type-7 (not depicted) do not have this property. CLS type-$7$ exhibit destructive interference on the same set of sites as CLS type-$3$ and were not included in the graphical representation of \cref{fig:quattro3gen} to avoid redundancy. The same reasoning applies to CLS type-$4$ and we only show CLS type-$6$ in the figure.

\subsubsection{Higher generations}
\begin{figure}[b]
	\centering
	\includegraphics[width=0.9\columnwidth]{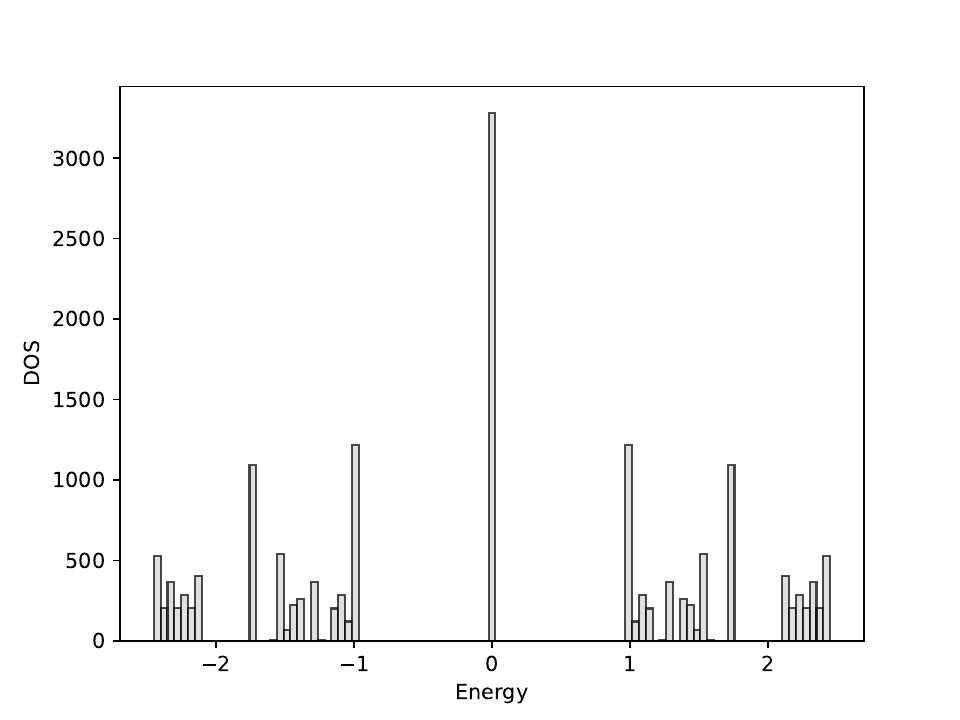}
	\caption{Hisotgram of the density of states in the $8$th generation of the fractal lattice. The energy interval is discretized with an energy step of $\Delta E=5\cdot 10^{-2}$.}
	\label{fig:dos}
\end{figure}

The number of CLS grows considerably at larger generations.
For the type-1 CLS---which is caused by the fractal's bipartiteness---, for instance, we can see that for the $n$-th generation, the number of sites in the two sublattices is given by $N_{\mathcal{A}}(n) = 6 \cdot 3^{n-1} - \sum_{i=1}^{n-1} 3^i$ and $N_{\mathcal{B}}(n) = 3^n$.
This results in
\begin{equation}
	\Delta N(n) = 6\cdot 3^{n-1} - \sum_{i=1}^{n} 3^i
	\label{eq:delta_n}
\end{equation}
zero-energy CLS, as predicted by Lieb's theorem. For instance, in the $8$th generation, there are $N=16404$ sites in total, with an imbalance in the two sublattices given by $N_\mathcal{A}=9843$ and $N_\mathcal{B}=6561$. The number of zero-energy CLS is thus $\Delta N = 3282$. The plot of density of states (DOS), in \cref{fig:dos}, provides a clear visualization of the zero modes present in our fractal lattice. An analysis of the DOS in the tight-binding regime on the Sierpinski lattice was done in Ref. \cite{Kempkes} to show the self-similarity at different scales.
For large $n$, the first term in \cref{eq:delta_n} becomes dominant, from which we conclude that the ratio $\alpha = \Delta N(n+1)/\Delta N(n)$ thus equals $3$.
Since the sides of the fractal are duplicated whenever going from one generation to the next,
the dimension is defined as
\[
d = \frac{\log{\alpha}}{\log{2}}
\]
and we thus observe that the dimensionality we assign to number of zero-energy CLS tends to the dimension of the fractal, $d_{H}=\log{3}/\log{2}$.

\begin{figure}[t]
	\centering
	\includegraphics[width=0.8\columnwidth]{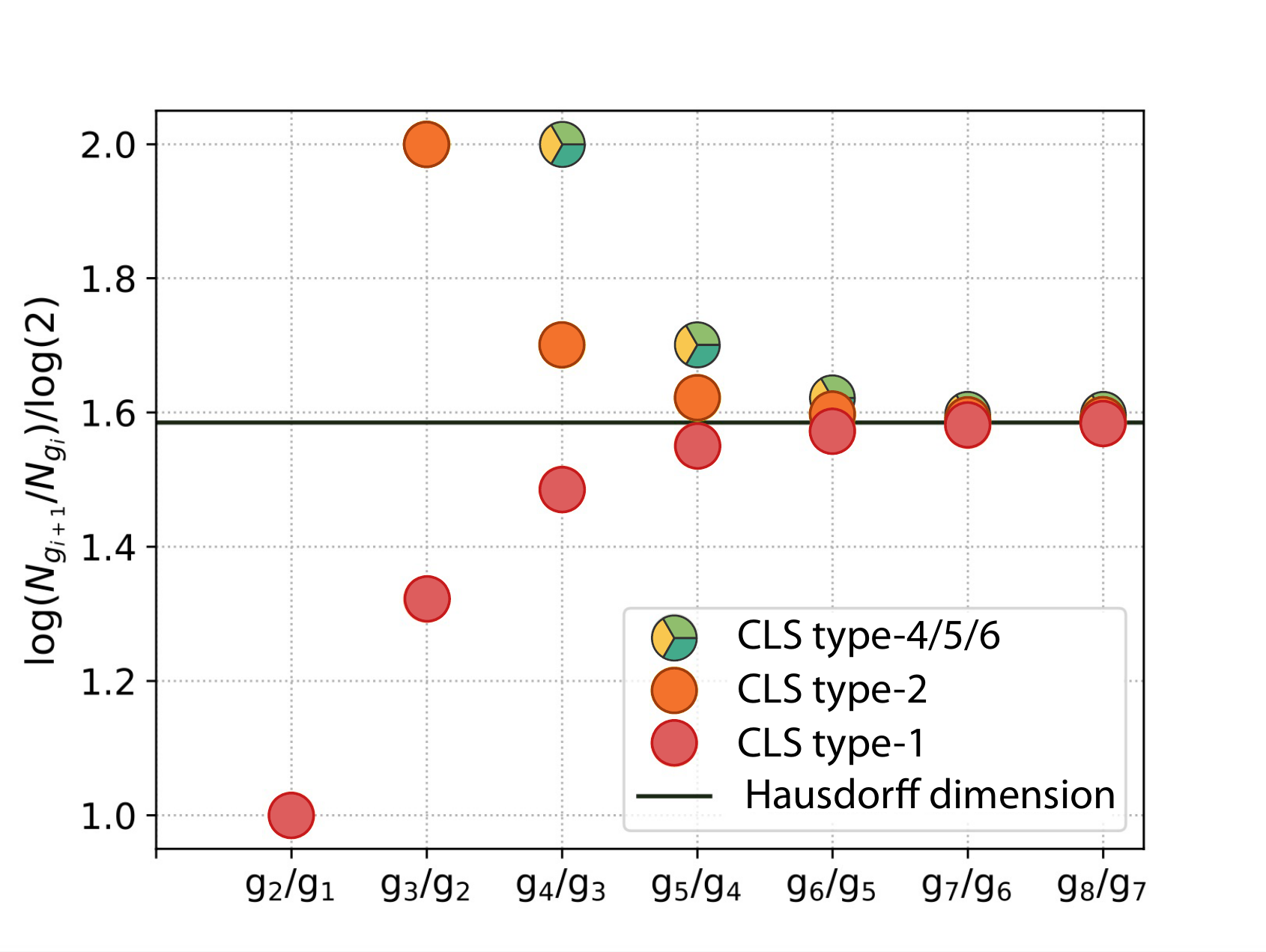}
	\caption{
		Scaling of the number $N_{g}$ of CLS of different types when increasing the generation $g$ of the Sierpinski triangle, from first generation $g_1$ to generation $g_8$.}
	\label{fig:scalingnumb}
\end{figure}

By explicitly diagonalizing the system for the first eight generations, we also studied how the number of the type-2 to type-7 CLS changes when considering higher generations of the lattice.
The results are presented in \cref{fig:scalingnumb}.
The number of CLS of types $1,2,4,5$ and $6$ tends to scale as the Hausdorff dimension when the generation number increases.

We notice that CLS of type-$4, 5$ and $6$ share the same scaling, even if they belong to different energy levels, see \cref{fig:energyspec3gen}. Regarding CLS of type-$3$ and type-$7$, we found that the number of states in the corresponding levels $\pm \sqrt{2}$ and $\pm \sqrt{5}$ does not increase but remains the same for odd generations, and is zero for even generations.

Motivated by these scaling behaviours, we also studied how the many-body ground-state energy $E_T$ around half-filling (i.e. the value of energy corresponding to the degenerate level in \cref{fig:lattice_energytot1gen}(b) changes when increasing the generation.
\begin{figure}[b]
	\centering
	\includegraphics[width=0.8\columnwidth]{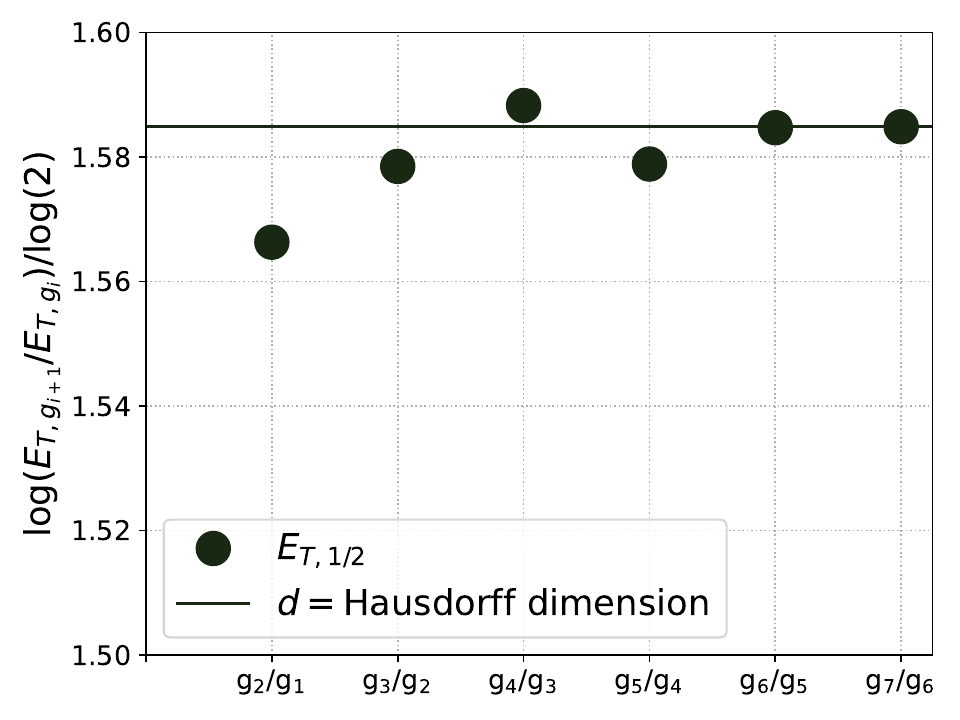}
	\caption{
		Scaling of the ground-state many-body energy at half-filling $E_{T,1/2}$ when increasing the generation $g$ of the Sierpinski triangle, from first generation $g_1$ to generation $g_8$.}
	\label{fig:scalingnenergy}
\end{figure}
As shown in \cref{fig:scalingnenergy}, these values also scale as the Hausdorff dimension. Therefore, they can be derived by recursive computations upon increasing the generation, without the need of diagonalizing the Hamiltonian. This also holds for the number of CLS in some energy levels, and constitutes a great advantage because the computational cost increases rapidly as we increase the generation.

As a last addition to the analysis of the zero-energy level, we investigated the average density on the corners in the last configuration of the zero-energy level of the many-body spectrum. This means, for example, filling $N_{\sigma}=6$ for the first generation, $N_{\sigma}=15$ for the second, and so on. We noticed that its value is $\bk{n}=1.5909$ for every generation that we implemented, very close to the Hausdorff dimension.

\subsection{Compact localized states when considering intrinsic spin-orbit coupling} \label{sec:ISOC}
So far, we have seen that CLS naturally appear in the Sierpinsky triangle when treated within a simple tight-binding formalism.
To emphasize the importance and universality of these states, we will demonstrate in the following that they are still present in the model even when taking intrinsic spin-orbit coupling (ISOC) into account.

To introduce ISOC into the model, we start from the TB Hamiltonian of the Sierpinsky triangle and add the term
\begin{equation} \label{eq:ISOC}
	H_{\text{SO}} = i b \sum_{\langle \langle i,j \rangle \rangle} v_{i,j} c^{\dagger}_{i} c_{j}
\end{equation}
to the Hamiltonian.
Here, the sum goes over pairs of sites that fulfil both of the following conditions: (i) they both are in the majority sublattice, and (ii) in the original Hamiltonian without ISOC, they are two sites apart from each other, i.e. the next-nearest neighbor (NNN). The coefficient $v_{i,j} = +1$ when the two-hop path from site $i$ to site $j$ is in the clockwise direction, and $v_{i,j} = -1$ when it is in anti-clockwise direction. The Hamiltonian includes intrinsic spin-orbit coupling (SOC), Rashba SOC, and a staggered mass in a honeycomb tight-binding lattice \cite{Kane2005PRL95146802Z2TopologicalOrder}. Since our focus is on generating topological features, we considered only the intrinsic SOC term in the Hamiltonian because this is the term that leads to the quantum spin Hall effect. The staggered mass opens a trivial gap and the Rashba SOC tends to close the topological gap. It is important to note that, because spin is a conserved quantum number, it can be omitted from the Hamiltonian during calculations. At the conclusion of the calculations, the spin can be reintroduced, with the understanding that all results are valid for the opposite spin as well, but with an inverted sign for edge propagation. 

\begin{figure}[H] 
	\centering
	\includegraphics[width=0.96\linewidth]{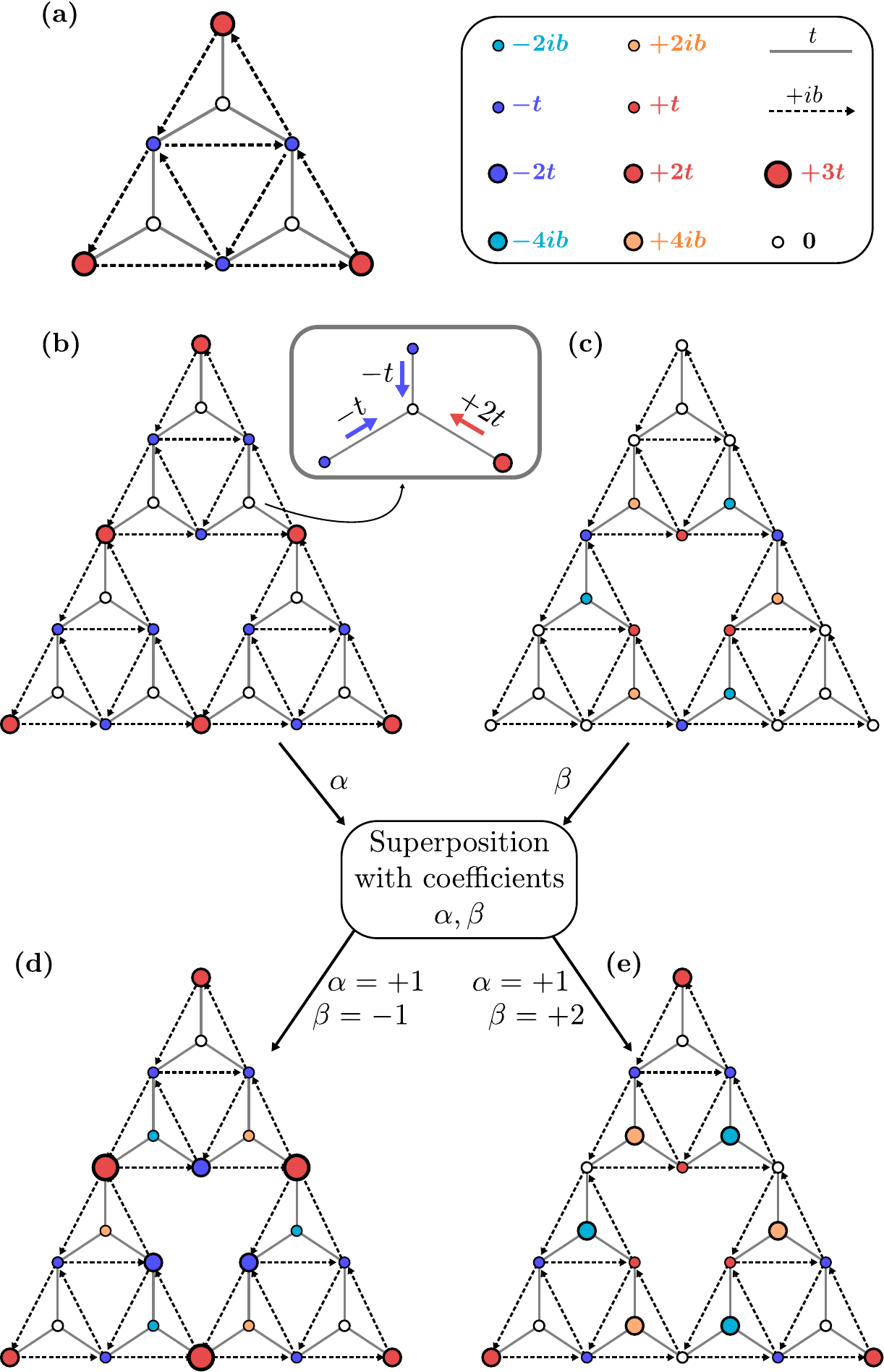}
	\caption{
 Graphical representation of both the Hamiltonian and specific eigenstates for the first and second generation Sierpinsky triangle with intrinsic spin-orbit coupling of strength $b$. Each circle represents a physical site, with lines denoting coupling between these sites; solid grey lines denote the normal coupling (with strength $t$), while the dashed lines denote the intrinsic spin-orbit coupling (which is complex and thus directional) of strength $i b$. The size and color of the circles encode the amplitude of the specific eigenstate shown.
(a) The only zero-energy eigenstate for the first generation of the Sierpinsky triangle with intrinsic spin-orbit coupling of strength $b$. (b) and (c) the two degenerate zero-energy eigenstates for the second generation. Superposing these, as shown in (d) and (e), gives a state with high enhancement at certain sites.
	}
	\label{fig:explainDegeneracy2}
\end{figure}

Effectively, \cref{eq:ISOC} adds complex-valued coupling (with amplitude $b$) between NNN sites---similar to Ref. \cite{vanGelderen2010PRB81125435RashbaIntrinsicSpinorbitInteractions}---, but only if the coupling does not cross the ``empty regions'' of the Sierpinski triangle.
In \cref{fig:explainDegeneracy2}(a), the Hamiltonian is shown in pictorial form, with complex couplings denoted by dashed, arrowed lines.

Let us now analyse the situation in detail. In contrast to the case without ISOC, the first few generations (we have checked until the fifth) only show one type of CLS, which lies at zero energy.
Moreover, in contrast to the case without ISOC, the CLS is non-degenerate in the first generation; the state is depicted in \cref{fig:explainDegeneracy2}(a).
Remarkably, it is exactly the state from \cref{fig:sumcorners}, so that the amplitudes are enhanced at the three outer corners of the fractal. The ISOC entangles the three degenerate eigenstates into a sum of the them.

In the second generation, the CLS become two-fold degenerate.
The first of the corresponding eigenstates is shown in \cref{fig:explainDegeneracy2}(b). 
We note that it is similar to the one from \cref{fig:explainDegeneracy2}(a), and indeed can be obtained from the latter by simply copying it three times.
In the following, we will call this eigenstate the \textit{ubiquitious state}. It sill vanishes on a large number of sites, which are marked in \cref{fig:explainDegeneracy2}(b) in black. 
Again, as with any CLS, the cause is destructive interference; see inset of \cref{fig:explainDegeneracy2}(b). 
The second eigenstate with zero energy is what we call a \textit{closed-loop state}. It loops around the central gap (white space) of the fractal and vanishes outside of it. 
We further note that the amplitudes of this closed-loop state depend on the coupling $t$ and the strength $b$ of the intrinsic spin-orbit coupling.

Similar to \cref{fig:sumcorners}(d), we can superpose the two degenerate zero-energy states to obtain a state which has a larger enhancement on some sites. This is demonstrated in \cref{fig:explainDegeneracy2} (d) and (e).
\begin{figure*}[h!] 
	\centering
	\includegraphics[width=0.8\linewidth]{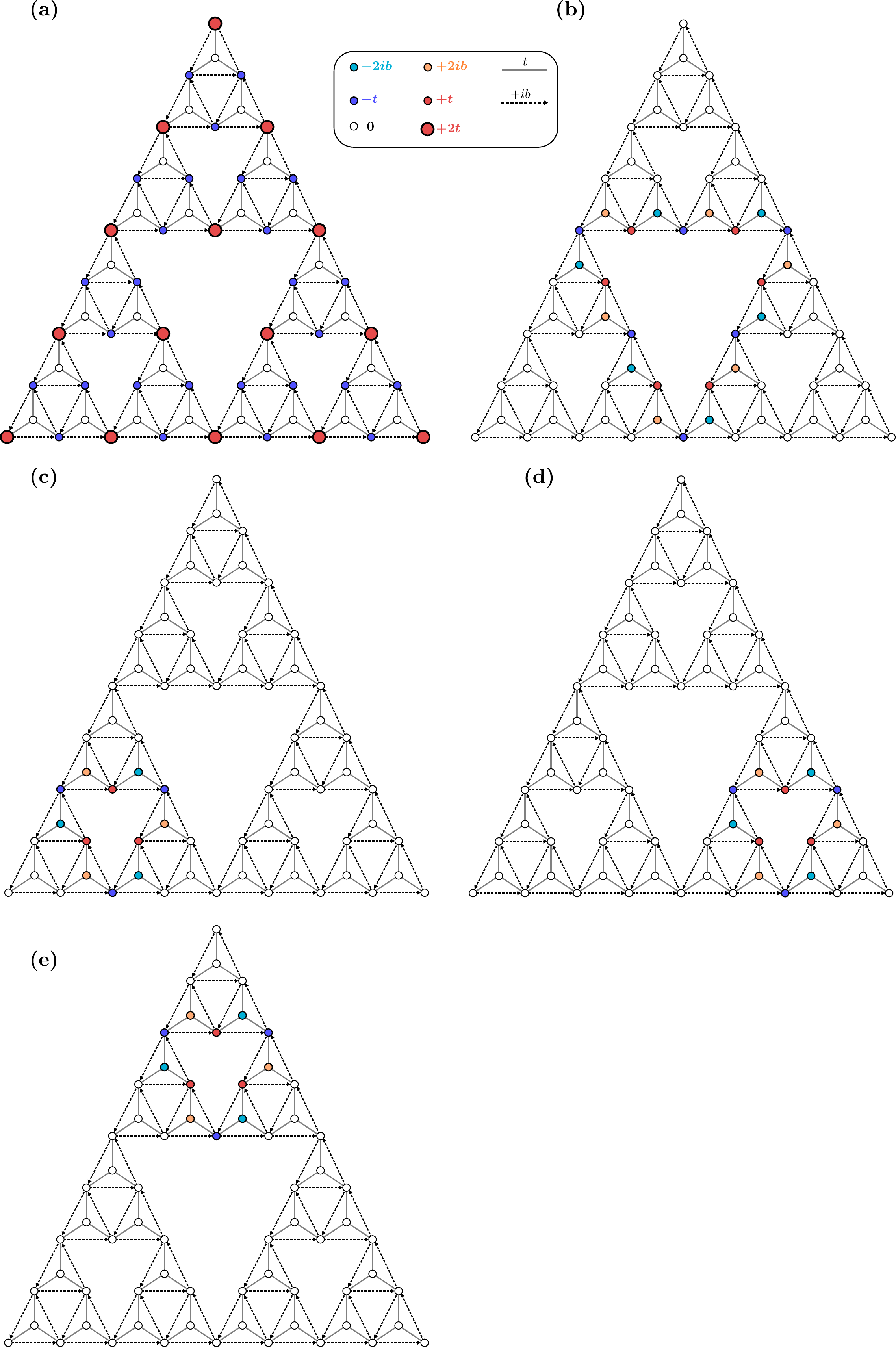}
	\caption{Graphical representation of both the Hamiltonian and specific eigenstates for the third generation Sierpinsky triangle with intrinsic spin-orbit coupling of strength $b$. Each circle represents a physical site, with lines denoting coupling between these sites; solid grey lines denote the normal coupling (with strength $t$), while the dashed lines denote the intrinsic spin-orbit coupling (which is complex and thus directional) of strength $i b$. The size and color of the circles encode the amplitude of the specific eigenstate shown. (a) to (e) show the five zero-energy eigenstates of the third-generation Sierpinsky triangle (see text for details).
	}
	\label{fig:explainDegeneracy3}
\end{figure*} 
In higher generations, the overall picture does not change much. For instance, in the third generation, there are five zero energy eigenstates: One ubiquitous (\cref{fig:explainDegeneracy3}(a)), one large closed-loop state for the central gap (\cref{fig:explainDegeneracy3}(b)) and three smaller closed-loop states for the smaller gaps (\cref{fig:explainDegeneracy3} (c), (d), and (e)).
Once again, one may superpose these states to obtain states with enhanced amplitudes on certain sites (corners, for instance).

We have checked the number $N$ of zero-energy CLS until the fifth generation. For each generation, this number is equal to one third of the number of zero-energy states without ISOC. Thus, the scaling of the number of such states is the same, with or without ISOC.

\section{\label{sec:QMC}The Hubbard Model}
We now introduce interaction to the model in the fractal lattice and present both the numerical methods used and the results of our implementations.

The full Hamiltonian in \cref{eq:hubb_ham} contains a two-body term, which prevents us from diagonalizing the system at a single-particle level, as done for the TB model. In the next subsections, we present three distinct methods that we use to investigate this Hamiltonian numerically.\\

\subsection{Exact Diagonalization\label{subsec:Exact}}

The first method consists in naively diagonalizing the Hamiltonian in a many-body basis of the Fock space and identifying the ground-state energy as the lowest energy eigenvalue and the ground-state wavefunction as its corresponding eigenvector. To diagonalize the Hamiltonian, we first need to choose a basis and our convention for the basis choice is, in the Fock space, strings with $2\times M$ entries that can take values of either $0$ or $1$ and that identify the positions of the electrons on the lattice, $1$ means that there is an electron, $0$ that there is none \cite{EDnew}.

\clearpage

We also need to define an order: the first half of the entries represents spin up particles populating the numbered lattice sites, while the second represents spin down particles with the same lattice site ordering. The number of such basis states is $4^{M}$. Since this number grows exponentially with the number of sites, exact diagonalization is a computationally demanding method to solve the Hamiltonian and it will only be used to solve small-size systems with a small number of electrons.

\subsection{Mean-Field Approximation\label{subsec:MFHF}}

One way of circumventing the two-body issue is to perform a mean-field Hartree-Fock approximation. This consists in rewriting the interaction term as \cite{Fazekas} 
\begin{multline}
	n_{i,\uparrow}n_{i,\downarrow} \simeq 
	\bk{n_{i,\uparrow}}n_{i,\downarrow} 
	+ \bk{n_{i,\downarrow}}n_{i,\uparrow} 
	- \bk{c^{\dagger}_{i,\uparrow}c_{i\downarrow}}c^{\dagger}_{i,\downarrow}c_{i\uparrow} \\
	- \bk{c^{\dagger}_{i,\downarrow}c_{i\uparrow}}c^{\dagger}_{i,\uparrow}c_{i\downarrow} 
	+ \bk{n_{i\uparrow}n_{i\downarrow}}
	+ \bk{c^{\dagger}_{i,\uparrow}c_{i\downarrow}}\bk{c^{\dagger}_{i,\downarrow}c_{i\uparrow}},
	\label{eq:interaction_MF}
\end{multline}
which allows us to express the Hubbard Hamiltonian in terms of one-body operators and proceed by constructing the many-body ground state in the same way as we did in the TB limit. The difference is that the mean-field term is not known, since the averages present in \cref{eq:interaction_MF} require the knowledge of the solution, which is, in turn, what we want to compute. The way we proceeded is by applying a self-consistent iterative scheme. Starting from a randomly generated Hamiltonian, we use the diagonalization approach introduced in the previous section to compute the many-body ground-state energy, average density per site, and wavefunction. The latter is then used to build a mean-field Hamiltonian $H_{MF}$, from which we again compute the above mentioned quantities. This scheme is iterated until the solutions converge. We show later that this approximation is valid for very small values of the interaction parameter $U$.

\subsection{CP-AFQMC}

The method that we largely implemented to study the Hubbard Hamiltonian is CP-AFQMC \cite{Nguyen, Zhang, Zhang_bosons, Zhang_symm}.
The starting point is a mean-field ansatz, found by following the procedure outlined in \cref{subsec:MFHF}. The imaginary-time evolution of this ansatz tends to the ground-state of the Hubbard Hamiltonian as the imaginary time becomes large. Implementing imaginary-time evolution requires a Hubbard-Stratonovich transformation, which introduces external auxiliary fields at each lattice site. This enables to go from an interacting system to a non-interacting system living in a space of fluctuating external fields. The wavefunction is written as a linear combination of many wavefunctions, the walkers. Evolving the full system in imaginary time means evolving each of the walkers.

In addition, more techniques are required to render the method efficient. Importance sampling is used, and the simulation is guided by the initial ansatz throughout the evolution in imaginary time. Back propagation is required to compute observables that do not commute with the Hamiltonian. A constrained path approximation  makes it possible to deal with the sign problem that stems from the fermionic nature of the particles populating the system. For a more detailed description of the method, we refer to \cref{app:cpmc}.

Our starting point was the validation of the QMC method by comparing its results with exact ones, i.e. with results found by performing exact diagonalization (ED). We also compared these two methods with the mean-field (MF) approach to understand to which extent this approximation is appropriate. Since the simulation time required by ED scales exponentially with the system size and the electron number, we considered a small system (the first generation of the fractal lattice), with a couple of electrons ($N_{\sigma}=1$). The quantities computed were the total, kinetic, and potential energy of the many-body ground-state, for values of $U\in[0.1,9]$. The potential energy is the contribution to the energy given by the Coulomb interaction in the Hubbard Hamiltonian in \cref{eq:hubb_ham}.
\begin{figure}[t]
	\centering
	\includegraphics[width=\columnwidth]{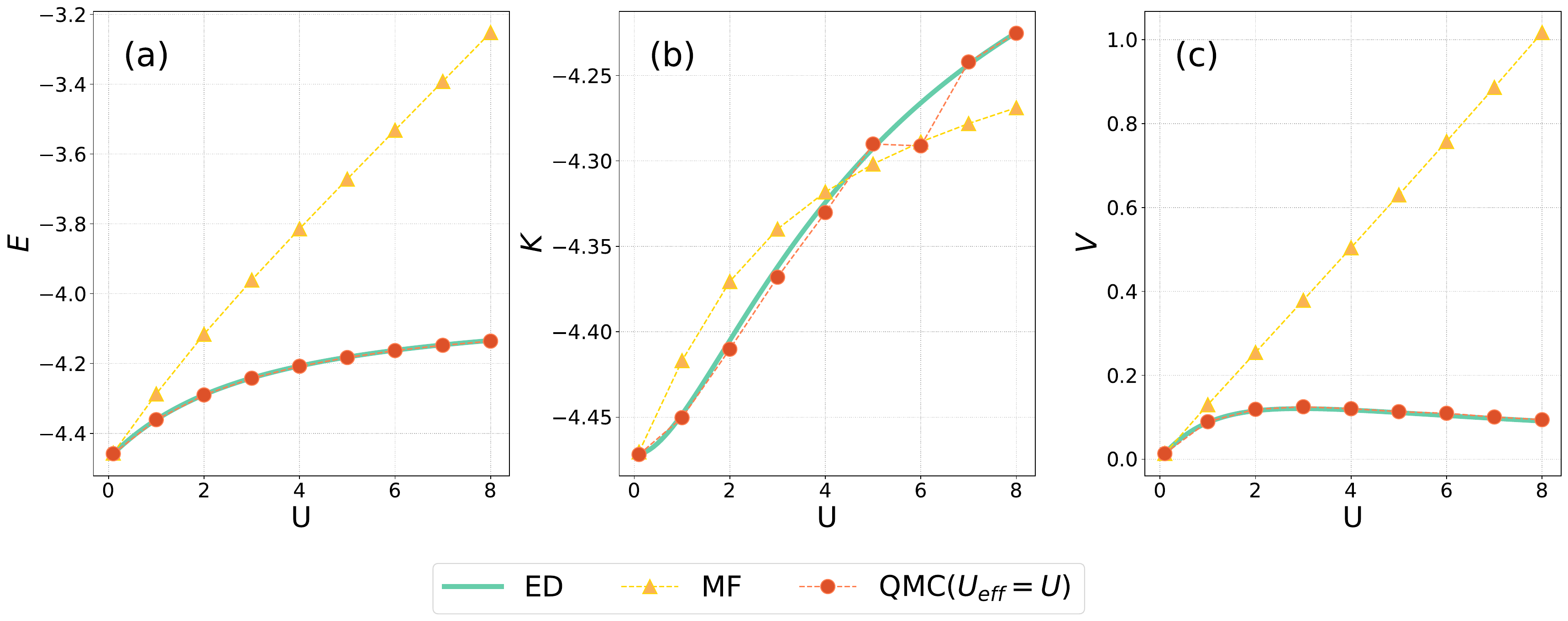}
	\caption{
		Comparison of the ground-state many-body energies as a function of interaction. We consider the first generation of the fractal lattice with $N_{\sigma}=1$. These energies are computed with three different implementation methods: ED (green line), MF (yellow triangles), and QMC (red dots). (a) The total energy, (b) the kinetic energy, and (c) the potential energy. The error bars associated to the QMC results are smaller than the pointer's size.}
	\label{fig:validity}
\end{figure} 
In \cref{fig:validity}, we present the behaviour of the energies computed using these three different methods, as the interaction strength increases. Strictly speaking, the mean-field (MF) approximation is valid only for $U=0$, but it is a reasonable approximation for very small values of the interaction U when $U \to 0$. On the other hand, the QMC approach follows very accurately the exact behaviour. For this reason, we focus on the QMC implementation and use MF only to make it more efficient. We also notice that the value of the kinetic energy computed by QMC for an interaction $U = 6$ deviates from the exact solution and coincides with the MF solution. This could be a consequence of the fact that the MF ansatz, used in an importance-sampling scheme and in the CP approximation, is biasing this result.

\begin{figure}[t]
	\centering
	\includegraphics[width=0.9\columnwidth]{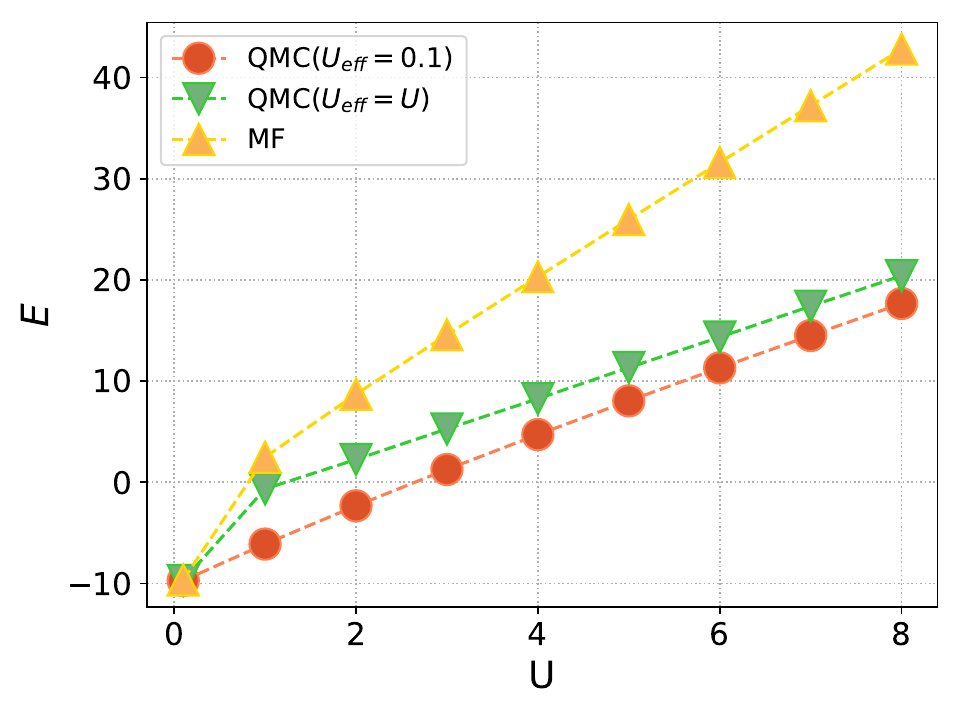}
	\caption{Comparison of the ground-state many-body energies as a function of interaction in the first generation of the fractal lattice with $N_{\sigma}=6$. These energies are computed with three different implementation methods: MF (yellow triangles), QMC ($U_{\text{eff}}=U$, red dots) and QMC ($U_{\text{eff}}=0.1$, green triangles). The error bars associated to the QMC results are smaller than the pointer's size.}
	\label{fig:validity2}
\end{figure}

To understand the extent of this influence, we performed simulations in the first generation with $N_{\sigma}=6$. We studied the behaviour of the QMC energies as the interaction strength is increased, with the simulation being guided by different MF ansätze. In particular, we performed simulations where the MF ansatz is generated using $U_{\text{eff}}=U$, where $U_{\text{eff}}$ is the interaction strength used in the MF simulation, referred to as effective potential. This behaviour is indicated by the inverted triangles in \cref{fig:validity2}. Then, we performed simulations with $U_{\text{eff}} = 0.1$. The simulation that provides the smallest total energy is the best approximation of the ground state, due to the variational principle. It is possible to observe that the latter ansatz gives better results for every value of interaction.
From this analysis, we can conclude that a preliminary study on the influence of the MF ansatz on the QMC simulation has to be performed before proceeding with the implementation. In fact, since the path of the walkers is influenced by the MF trial wavefunction, a bad ansatz affects negatively the paths, making them deviate from the exact solution. Moreover, to handle the sign problem, we introduce a constrained-path approximation that renders systematic errors. However, those errors are small since the results coincides very closely with exact diagonalization, as shown in \cref{fig:validity}.

\subsection{Interaction and density distribution}

We now consider the configuration where $N_{\sigma}=6$, which corresponds to the last energy state in the degenerate energy-level found when $U=0$, see \cref{fig:lattice_energytot1gen}(b). In the TB studies, we understood that this state is a consequence of the CLS, with destructive interference happening at the sites in the centre of the triangles. We also observed that the density at the corners has a value close to the Hausdorff dimension. Studying this configuration allows us to tackle the consequences of interaction in this type of states. 

\begin{figure}[t]
	\centering
	\includegraphics[width=0.9\columnwidth]{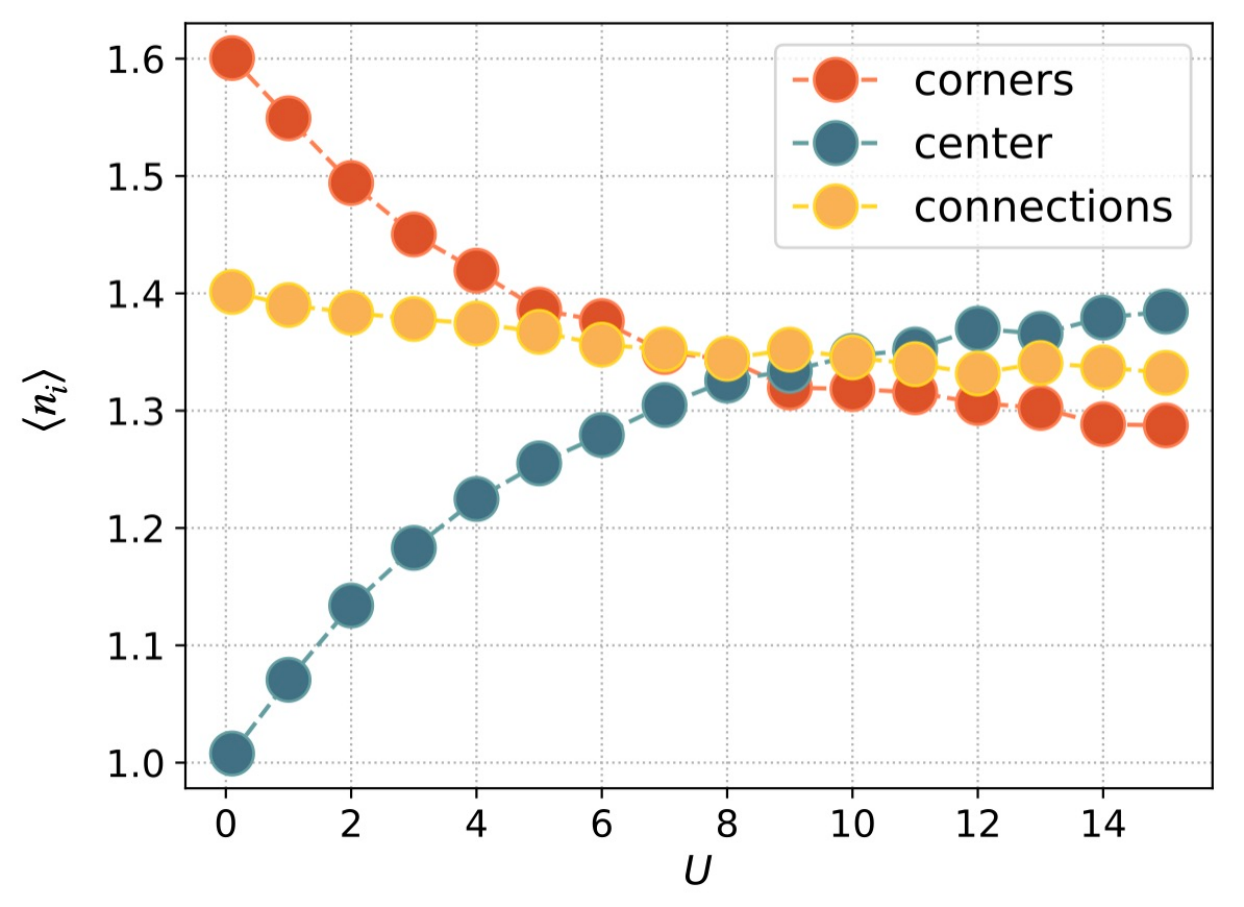}
	\caption{Average density per site $\bk{n_i}$ as a function of interaction strength $U$ in the three different groups of states, on the first generation of the fractal lattice with $N_{\sigma}=6$.}
	\label{fig:eff_interact}
\end{figure}

We can divide the lattice into three groups of sites: sites $1,6,9$ called \textit{corners}, sites $2,5,7$ called \textit{center} and sites $3,4,8$ called \textit{connections}, see \cref{fig:lattice_energytot1gen}(a) for the site indexing. We expect the density to be the same in each of these groups, since there is no reason for an imbalance. \Cref{fig:eff_interact} shows how the average density per site $\langle n_{i}\rangle$ changes in these groups of sites when increasing the interaction strength.
We observe that for small interactions, both corners and connection sites are more populated than the center sites. As the interaction increases, their density decreases, while it grows in the sites at the center. The electrons start spreading towards the center of the lattice, and they keep spreading until the density is approximately homogeneous on every site, at a value of approximately $\bk{n_i}\sim 1.35$ and $U\sim 8$. Thereafter, the density in the center sites continues to increase, while decreasing slowly on the rest of the lattice sites, meaning that electrons start to accumulate in the center sites. For very strong interaction values, it becomes more favourable to have electrons in sites with more connectivity, where hopping is more probable. 

Relating this study to the TB considerations and the CLS type-1 discussed in \cref{sec:CLS}, we expect that those types of states with destructive interference on center sites get destroyed as soon as the interaction is turned on, since the density on those sites increases.

\subsection{Quantum phases in the second generation of the fractal lattice}

The quantum phases of the Hubbard Model have been intensively studied on various lattices, in particular for the case of half-filling. We consider here the second generation of the fractal lattice, \cref{fig:fractal_lattice}, which has $24$ sites, at half-filling, $N_{\sigma}=12$. 

\begin{figure}[t]
	\centering
	\includegraphics[width=\columnwidth]{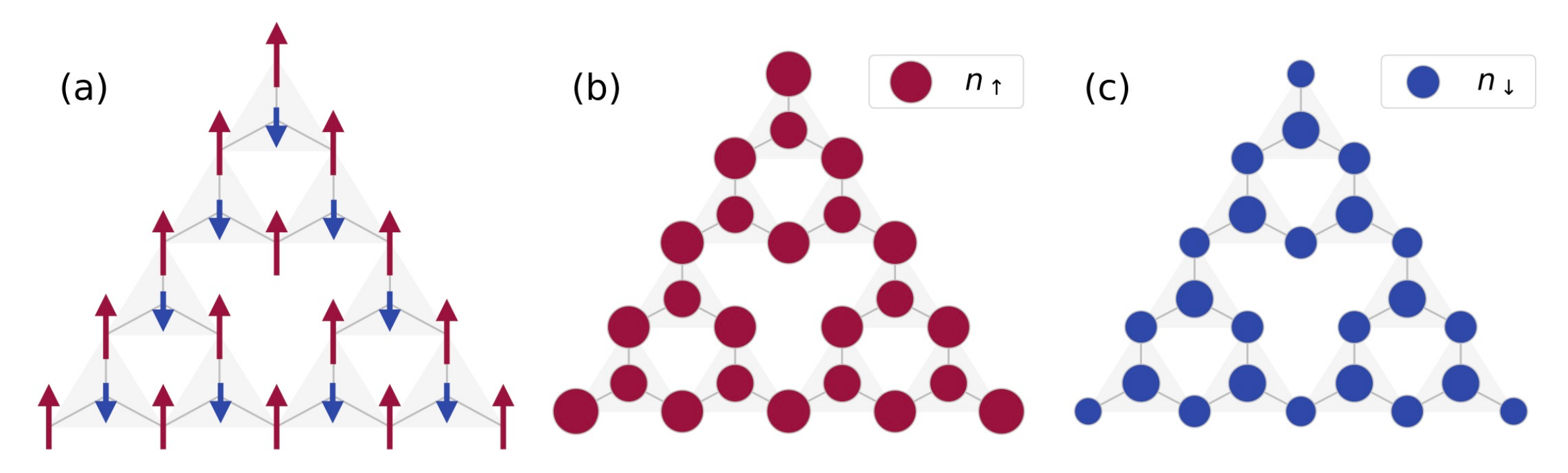}
	\caption{Local magnetization and average density at half-filling on the second generation with interaction parameter $U=0.1$. (a) Local magnetization per site, it shows a ferrimagnetic state; red (blue) arrows pointing up (down) indicate positive (negative) local magnetization. The length of the arrows represents the intensity of the local magnetization. (b) Average density per site of spin-up electrons $n_{\uparrow}$. (c) Average density per site of spin-down electrons $n_{\downarrow}$. The magnitude of the dots is representative of the intensity of the density.}
	\label{fig:ferrimagnet}
\end{figure}

Let us start by studying the behaviour of the magnetisation. We define the local magnetization $m_i$ and the total magnetization $m_\alpha$ per sublattice as
\[
m_i = \bk{n_{i,\uparrow}} - \bk{n_{i,\downarrow}}, 
\]
\[
m_{\alpha} = \frac{1}{N_{\alpha}} \sum_{i \in {\Lambda}_{\alpha} } m_{i},
\]
where the index $\alpha$ refers to one particular sublattice, $\alpha= \mathcal{A}, \mathcal{B}$, and ${\Lambda}_{\alpha}$ refers to the set of site indices belonging to the $\alpha$ sublattice. \cref{fig:ferrimagnet}(a) shows the magnitude of the local magnetization for weak interaction on the lattice structure. We see that the magnetization on different sublattices has opposite sign, and the magnitude in one is larger than the magnitude in the other, thus characterizing a \textit{ferrimagnetic} phase. A mean-field study of the Hubbard model in fractal-honeycomb lattices also finds spontaneous spin polarization~\cite{Pedersen2020}, which further corroborates our observation of a ferrimagnetic state.

The local magnetization is related to the projection of the spin along a certain axis. Since the sum of local magnetizations on the lattice is clearly not zero, we find that the system equilibrates to a spin imbalanced configuration. In particular, we find that $N_{\mathcal{A}}=15$ and $N_{\mathcal{B}}=9$. Without loss of generality, we investigate the case with $N_{\uparrow}=N_{\mathcal{A}}$ and $N_{\downarrow}=N_{\mathcal{B}}$. The opposite case is obtained by inverting the positive direction along the axis where the spin is projected. Figs. \ref{fig:ferrimagnet}(b) and (c) show the distribution of the $15$ spin-up and the $9$ spin-down electrons on the lattice, in the weak interacting regime. We can interpret \cref{fig:ferrimagnet}(b) by considering that the $9$ spin-up and spin-down electrons fill the single-particle energy levels with negative energy, see \cref{fig:energyspec2gen}. The remaining $6$ spin-up electrons are placed in the degenerate zero-energy level.
In this level, the CLS have destructive interference on the sites in the centre ($\mathcal{B}$ sublattice); sites on sublattice $\mathcal{A}$ get more densely populated. Therefore, the origin of imbalanced magnetic properties at half-filling and weak interaction can be reconnected to the zero-energy CLS that we found at a TB level, which also vanishes at the $\mathcal{B}$ sublattice.
\begin{figure}[t]
	\centering
	\includegraphics[width=0.9\columnwidth]{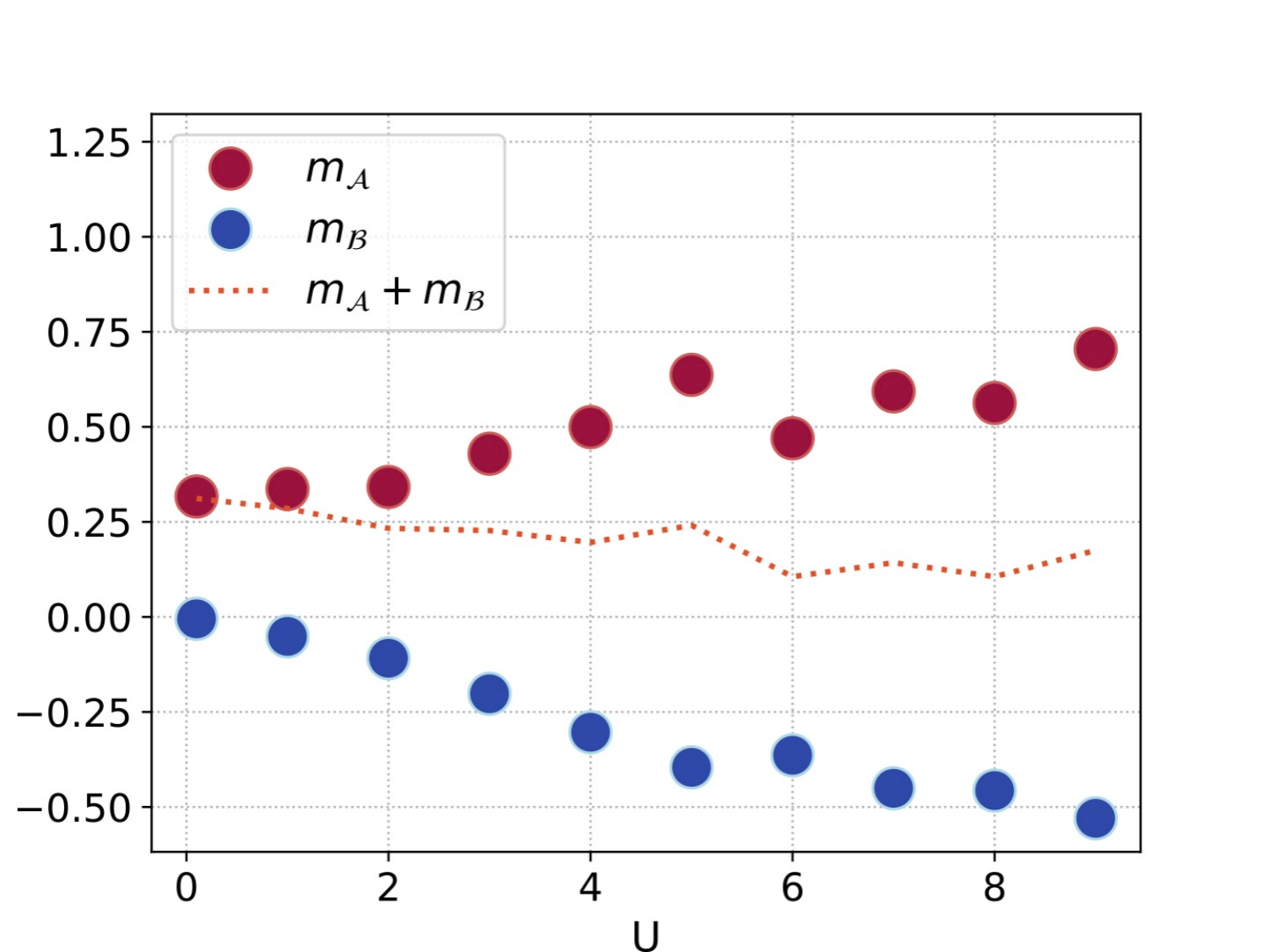}
	\caption{Average of the local magnetization on the two sublattices for different values of interaction parameter $U$.}
	\label{fig:local_magn}
\end{figure}

Now, we want to investigate the behaviour of the system when increasing the interaction strength. First, we notice that the imbalance between spin-up and spin-down electrons remains when increasing the interaction strength. Regarding the local magnetization, in \cref{fig:local_magn} we show the total magnetization in the two sublattices as a function of interaction strength. We observe that the sum of the magnetization in the two sublattices almost does not change. This quantity represents the total magnetization along a projection axis, and since the ratio of number of spin-up and spin-down electrons remains unchanged, we expect a constant behaviour. However, we need to take into account the fact that the number of spin-up and spin-down electrons is an output of the QMC simulation and, thus, subject to subtle fluctuations. For a regular periodic lattice, the theory predicts antiferromagnetic behaviour in the strong coupling regime, where the average local magnetization per site in both sub-lattices is $m_{\alpha}$. The reason why we are not able to see this behaviour is due to the imbalance in the number of electrons that populated the system which, in turn, is a consequence of the geometry of the lattice. 

\begin{figure}[t]
	\centering
	\includegraphics[width=0.9\columnwidth]{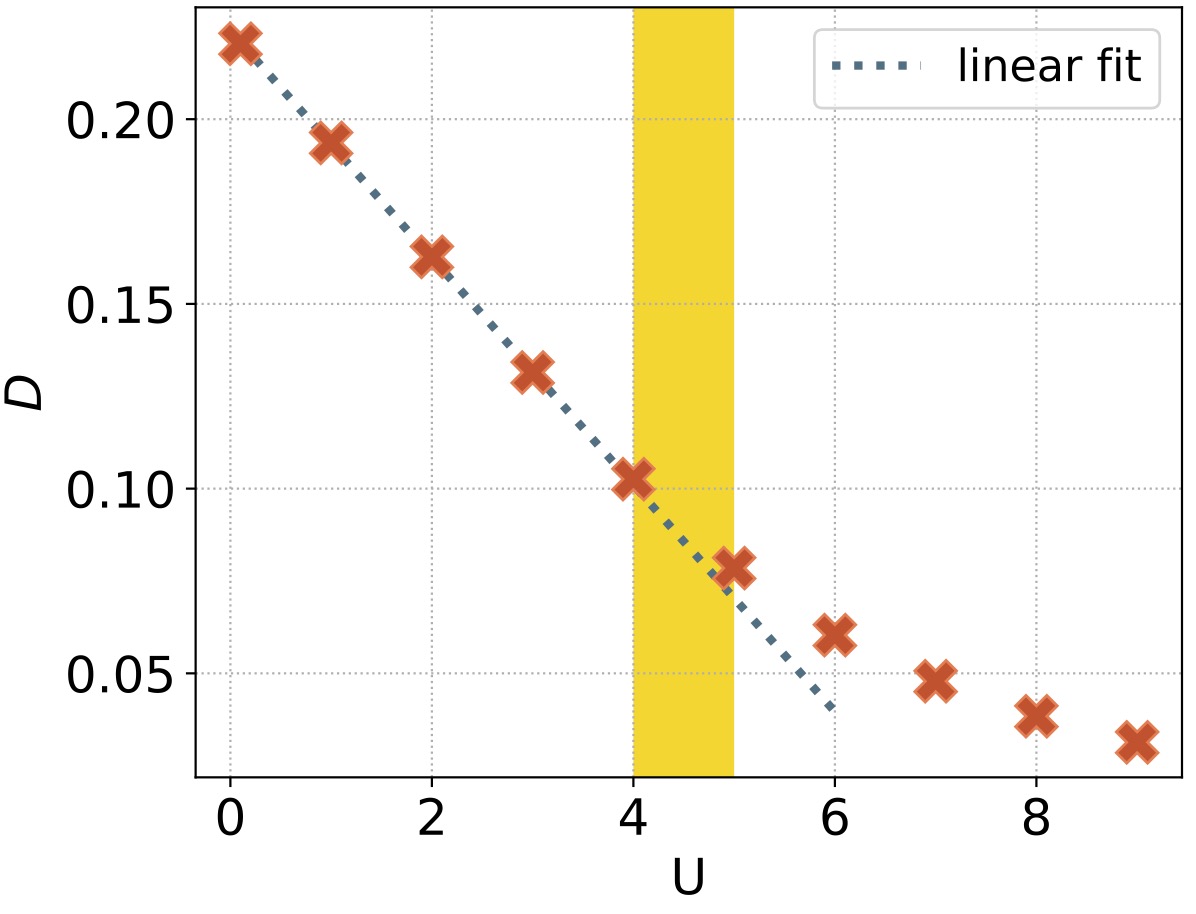}
	\caption{Doublon density as a function of the interaction parameter $U$. The behaviour is linear until a critical value of $U_C \sim 4.5$.}
	\label{fig:doublon}
\end{figure}

Another quantity that can be studied to determine the phases of the system is the doublon density, defined as
\[
D=\frac{1}{N}\sum_{i} \bk{n_{i\uparrow}n_{i\downarrow}}
\]
where the sum runs over the lattice sites. While being easy to compute, the double occupancy has the advantage of being related to the metallic behaviour of the system. Moreover, studies on periodic lattices suggest that the behaviour of $D$ as a function of interaction should differentiate metallic and insulating phases \cite{Vinicius, honeycomb_useful}. In particular, Brinkman and Rice obtained, through a variational calculation, that the doublon density behaves linearly in the metallic phase \cite{Birkam_Rice}. Even though this result was derived in the context of conventional band theory, we also observe in \cref{fig:doublon} a linear behaviour of $D$ until a critical value of approximately $U_{c} \sim 4.5$. Curiously, the value of the critical interaction is the same as the one describing a Mott transition from a paramagnetic conducting state at small values of $U$, to an antiferromagnetic (AFM) insulator at $U>4.5$ in the $2$D honeycomb lattice \cite{Sorella}. After this value of interaction strength, the behaviour deviates from the linear one.

To further investigate the magnetic nature of the system, we consider the magnetic correlation function
\begin{equation}
	c_{m}(i,j) = \bk{m_{i}m_{j}},
	\label{eq:magn_corr}
\end{equation}
which quantifies the correlation between the local magnetization of a pair of electrons, one placed at site $i$ and one at site $j$.

\begin{figure}[t]
	\centering
	\includegraphics[width=\columnwidth]{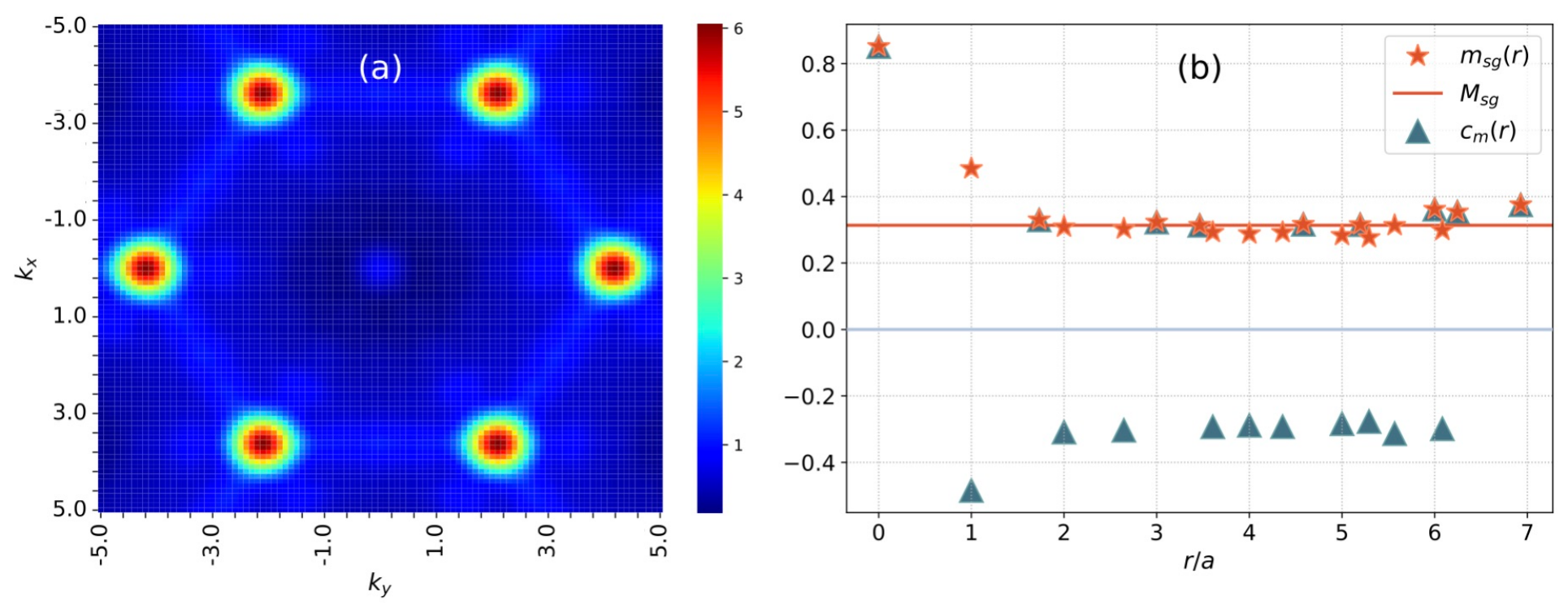}
	\caption{(a) Magnetic structure factor in the $\bm{k}=(k_{x}, k_{y})$ space for $U=6$. (b) Radial staggered magnetization (orange stars), radial magnetic correlation (blue triangles) and value of total staggered magnetization (orange horizontal line) for $U=6$. The horizontal light-blue line highlights the position of the origin along the $y$ axis. The radial distance $r$ is normalized by $a$, the distance between two neighbouring sites in the lattice. }
	\label{fig:struct}
\end{figure}

From the magnetic correlation, one can compute the magnetic structure factor, defined as
\[
S(\bm{k}) = \frac{1}{N^2} \sum_{i,j} e^{i \bm{k} \cdot \bm{r}_{ij}} c_{m}(i,j),
\]
where $\bm{r}_{ij} = \bm{r}_{i} - \bm{r}_{j}$. The symbol $\bm{k}$ indicates the momenta in the reciprocal space. It is possible to define a similar quantity, where instead of considering magnetic correlations, one considers density correlations. The computation of magnetic and density structure factors is a common method used to investigate both the metallic and spin phases of the system. 
The magnetic structure factor is used to detect magnetic order since it shows peaks at the K points of the Brillouin zone \cite{Vinicius}.     This method can be used here because it does not require knowledge of allowed momenta in the reciprocal space, which is not well defined for fractals. In \cref{fig:struct} (a), we show the magnetic structure factor in the strong interaction regime, where one observes the formation of peaks, which are a signature of magnetic order. We then computed the magnetic structure factor for various values of interaction and found a similar structure for each value. This seems to suggest that the system has spin order for any value of interaction strength.
Before continuing, let us mention that the six-fold rotational symmetry visible in \cref{fig:struct} (a) can be explained using the symmetries of the setup; see \cref{app:structureFactor}.

\begin{figure}[h]
	\centering
	\includegraphics[width=0.9\columnwidth]{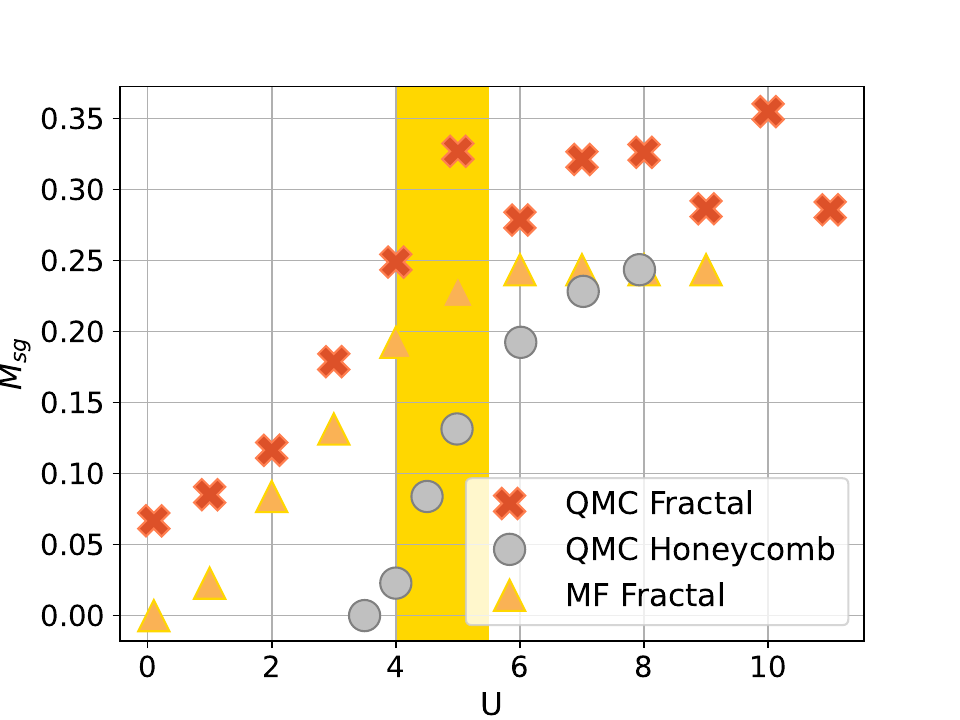}
	\caption{Total staggered magnetization as a function of the interaction strength $U$. For the second generation of the Sierpinski fractal, the calculations were performed using MF (yellow triangles) and QMC (orange crosses). For the regular honeycomb lattice, we extracted the QMC data from Ref. \cite{Raczkowski2020} (gray circles).}
	\label{fig:2gen_stgg}
\end{figure}

To validate the presence of magnetic order, we look at the radial behaviour of the magnetic correlation defined in \cref{eq:magn_corr}. In particular, we define the angle-averaged radial magnetic correlation,
\[
c_{m}(r) = \frac{1}{n_{r}}\sum_{\left\{ i,j \right\} \in n_{r}}c_{m}(i,j),
\]
where $n_{r}$ is the number of pairs $\left\{ i,j \right\}$ that have the same distance $r$. This quantity averages the magntic correlation functions of pairs of electrons located at sites $r$ far from each other. The set of possible distances $r$ is discrete, and it is important to notice that, for a given distance, all the pairs connect sites belonging either to the same sublattice or to different sublattices.  In \cref{fig:struct}(a), we show the dependence of this correlation on the radial distance $r$. The behaviour shows positive and negative correlations, which indeed can indicate antiferromagnetic or ferrimagnetic order. One of the properties of anti and ferrimagnetism is spin orientations that alternate in the lattice. This means that electrons on sites belonging to the same sublattice should have spins oriented to the same direction (the correlation is positive) while electrons on sites belonging to different sublattices should have spins pointing in opposite direction (the correlation is negative). To verify this behaviour, we multiply the spin radial correlation by a factor $(-1)^{\alpha}$, where $\alpha$ is even (odd) if sites $i,j$ belong to the same (different) sublattice. The quantity obtained will be referred to as radial staggered magnetization,
\[
m_{sg}(r) = \frac{1}{n_{r}}\sum_{\left\{ i,j \right\} \in n_{r}}  (-1)^{\alpha} c_{m}(i,j).
\]
The result is shown in \cref{fig:struct}(b). We can conclude that the property of opposite spins in different sublattices is satisfied, as expected from the analysis of the local magnetization for the different lattice sites. 
In order to study the magnetic order at different values of interaction, we investigate the total staggered magnetization,
\[
M_{sg} = \frac{1}{N_{r}} \sum_{i_{r}} m_{sg}(r),
\]
where the sum is over the number $i_{r}$ of possible distances $N_{r}$. This quantity represents the average staggered magnetization over different values of radial distances. Its value for $U=6$ is plotted in \cref{fig:struct}(b) as a reference to the eye. Its behaviour as a function of interaction is shown in \cref{fig:2gen_stgg}, where we see that the value increases with the interaction before the critical value around $U_{c}\sim 4.5$. After it seems to oscillate without an overall growth or reduction for larger $U$. The fact that it always assumes a vanishing value for every value of interaction proves the existence and resilience of magnetic order. In \cref{fig:2gen_stgg}, we also show the staggered magnetization obtained with QMC calculations in a recent work on a honeycomb lattice \cite{Raczkowski2020}, the regular version of our fractal lattice. In the regular case, the phase before the transition is not magnetic, as the total staggered magnetization is zero. Finally, the yellow triangles in \cref{fig:2gen_stgg} show the total staggered magnetization computed using the MF approach under the same conditions. The comparison with MF results at half-filling underlines the necessity of powerful methods such as QMC. In fact, QMC simulations equilibrate to a steady state with an imbalanced number of spin-up and spin-down electrons, even when initializing the calculation with balanced trial wavefunctions. This results in a finite magnetic order, also at weak interactions. In contrast, MF calculations at half-filling preserve the number of spin-up and spin-down electrons balanced, giving a non-magnetic state at $U=0$. As expected, the MF result shows a spurious transition to a magnetic phase.

Overall, we observe the formation of a ferrimagnetic order, which can be related, at weak interaction, to the zero-energy CLS states unveiled at a TB level and, in general, to the imbalance in the number of spin-up and spin-down electrons. The latter is a consequence of the geometry of the lattice. Therefore, we observe a phase transition from metallic to insulating phase, suggested by the change in behaviour of the doublon density and the maximum of the magnetic correlation function in $\boldsymbol{k}$ space. Since this phase transition is driven by interaction between electrons, it is a Mott transition. 

\section{\label{sec:Concl}Conclusion}
In this work, we solved the Hubbard Model for fermions on a fractal lattice, using various methods. 

Initially, we examined the non-interacting limit of the model, employing a TB approach. It successfully confirmed the presence of particle-hole symmetry through the examination of the average density distribution on the lattice and the symmetry of the total many-body energy. Additionally, by analyzing the energy spectrum of the Hamiltonian at the single-electron level, we identified the emergence of CLS at various energy levels. In the first generation, we observed the formation of a specific type of CLS, characterized by zero energy and destructive interference on sites with a connectivity of 3. Moving to the second generation, we witnessed the appearance of an additional type of CLS, manifesting at energy levels of $t$ and $-t$, and displaying destructive interference along the reflection axes. Progressing to the third generation, we observed the emergence of more diverse types of CLS, forming at new energy values and exhibiting destructive interference on other sublattices. Remarkably, the number of CLS at each energy level increased in higher generations of the fractal, scaling with the Hausdorff dimension of the fractal $d_{H}=\log{3}/\log{2}\simeq 1.58$. Moreover, we identified CLS of type-$3$ and $7$, emerging on energy levels $\pm \sqrt{2}$ and $\pm \sqrt{5}$, respectively. The number of these two types of CLS remains the same for odd generations and is zero for even generations.
Finally, we discovered that the density at the corners, for configurations with a number of electrons that fills the degenerate level at zero energy, closely approximates the dimension of the fractal across all generations of the lattice. This result resembles the findings of higher-order topological insulators realized using acoustic quantum simulators. Indeed, the outer corner modes were found to exhibit the same dimension of the Sierpinksi carpet\cite{Floq_exp}. Unlike in their framework, this interesting connection between outer modes and fractal dimension arises in the absence of any external magnetic flux. We note that CLS in Sierpinski fractals have been previously considered, both theoretically \cite{Pal2018PRB97195101FlatBandsFractallikeGeometry,Biswas2023PELSaN153115762DesignerQuantumStatesFractal,Nandy2021PS96045802ControlledImprisonmentWavePacket} and experimentally \cite{Xie2021AP6116104FractallikePhotonicLatticesLocalized,Hanafi2022AOM102102523LocalizedStatesEmergingSingular}. In addition, superconductivity \cite{Westerhout} and plasmonic properties \cite{Iliasov} were investigated in fractal structures. However, these works use a different arrangement and number of sites, since they do not consider the central site in each small filled triangle [cf. Fig. 2(a)]. Their results are thus not directly applicable to the setup considered here. While constructing fractal structures, one may consider sites at the vertices of the fractal, sites at the center of the ``bulk" parts of the fractal, or both. The first is the usual ``fractal lattice" construction and the second leads to the dual structure, which has the same Hausdorff dimension, but a different distribution of voids. A combination of both cases is considered here. Although a direct comparison with Refs \cite{Pal2018PRB97195101FlatBandsFractallikeGeometry,Biswas2023PELSaN153115762DesignerQuantumStatesFractal,Nandy2021PS96045802ControlledImprisonmentWavePacket, Xie2021AP6116104FractallikePhotonicLatticesLocalized,Hanafi2022AOM102102523LocalizedStatesEmergingSingular} is not possible, previous studies on fractal lattices showcase the relevance of fractal geometries, demonstrating, for example, their ability to host anyonic excitations \cite{Manna, Jaworowski}. We investigated the model in the presence of intrinsic spin-orbit coupling. We found that the zero-energy eigenstates become entangled and lead to high-intensity corner modes in the inner and outer triangles.

Subsequently, we numerically implemented the full model, including the interaction term, using three distinct numerical simulations. The first approach involved exact diagonalization, which is an exact method but limited in computational scalability. Therefore, we employed it primarily for validating the results obtained from the other two numerical methods, specifically for smaller system sizes and lower electronic fillings. The second approach entailed employing a mean-field Hartree-Fock approximation, which offers analytical insights but is observed to be valid only for relatively weak interaction strengths. The primary method, extensively employed in this study, is CP-AFQMC. 

When examining the effects of interaction, the application of CP-AFQMC enabled a comprehensive exploration of the density configurations. By concentrating on the electronic filling value that displayed density patterns closely aligned with the dimension of the fractal at the corners, we observed notable changes in the density distribution upon introducing interaction. We demonstrated that under strong interaction, sites with higher connectivity, where destructive interference occurred under weak interaction, became more populated as a countermeasure against the effects of interaction. We observe a decrease in the density at the corners as soon as interaction is introduced, leading to the conclusion that the CLS of type-1 are not robust in the presence of interaction. 

We further employed CP-AFQMC to investigate the quantum phases of the system within the second generation of the lattice at half filling, considering non-zero interaction strengths. Initially, we observed that the simulation reached an equilibrium state characterized by an unequal number of spin-up and spin-down electrons, corresponding to the sizes of the two sublattices. Our subsequent focus shifted towards studying the magnetic order within the system. To accomplish this, we computed the local magnetization per site and discovered that neighbouring sites exhibited opposite signs and varying magnitudes. This was consistently observed across all interaction strengths and indicated the presence of a ferrimagnetic phase, see \cref{fig:ferrimagnet}(a). We were able to establish a connection between the imbalance in local magnetization and the type-1 CLS under weak interaction. Subsequently, we verified the magnetic order by observing prominent peaks in the magnetic structure factor and long-range order in the magnetic correlations. Lastly, we detected a Mott transition by analyzing the behaviour of the tail of the magnetic correlations and the doublon density. This transition occurs at an approximate value of $U_{C} \simeq 4.5$, which intriguingly corresponds to the critical value detected through QMC methods for a Mott transition on a honeycomb lattice \cite{Sorella}, the periodic lattice configuration most similar to our chosen fractal structure. The difference is in the magnetic order, which we find to be ferrimagnetic for every value of interaction strength. The latter result, in turn, agrees with studies of Hubbard model on non-periodic lattices, such as the two-dimensional hexagonal golden-mean tiling \cite{Ferri}.

Concerning the effect of spatial dimensionality in the phase diagram of the ground state of the Hubbard model at half-filling, important differences are observed. In one dimension (1D), no Mott transition at finite interaction is predicted, with the system being metallic only for $U = 0$ and an AFM insulator for $U > 0$~\cite{Lieb1968}. Moreover, the low-energy excitations of the 1D Hubbard model bear resemblance to Tomonaga-Luttinger liquid theory~\cite{Haldane1980,Haldane1981,Haldane1981-2}. In 2D, the aforementioned phase diagram changes depending on the geometry considered. For the honeycomb lattice, which is the most similar to the fractal lattice considered in this work, a Mott transition is predicted at finite interaction~\cite{Raczkowski2020}. Below the critical interaction $U_C$, the system describes a non-magnetic semimetal, whereas above $U_C$, the honeycomb-Hubbard model describes an AFM insulator. Our calculations consider a spatial dimension that lies between 1 and 2, and new physics was observed. Similar to the 2D honeycomb model, a Mott transition at $U_C \simeq 4.5$ was observed, but no AFM state was found in the range of interactions considered. Instead, we observed that the system is always ferrimagnetic: a ferrimagnetic metal below $U_C$ and a ferrimagnetic insulator above it.
Therefore, our work shows that new physics emerges at fractal dimensions, which is different from the lower and upper integer boundaries. It remains to verify whether this behavior is specific to the Sierpinski triangle, with dimension 1.58, or whether it is generic to other fractal lattices with dimension between one and two. We hope that these findings will stimulate both theoretical and experimental research in interacting systems at non-integer dimension.

\section*{\label{sec:Akn}Acknowledgements}
We thank Lumen Eek for the useful insights on latent symmetries, which were sources of inspiration for part of this project. We are also grateful to T.S. do Espirito Santo for insightful discussions on the numerical methods used. C.M.S. acknowledges the research program “Materials for the Quantum Age” (QuMat) for financial support. V. Z. acknowledges financial support from the Brazilian agencies Coordenação de Aperfeiçoamento de Pesquisa de Pessoal de Nível Superior (CAPES) under the Netherlands Universities Foundation for International Cooperation (NUFFIC) exchange program (process number 88887.649143/2021-00) and the Serrapilheira Institute (grant number Serra-1812-27802). The QMC simulations were performed using a modification of release CP-AFQMC-v1.0 of the code available at the \href{https://github.com/quantechsimulations/CP-AFQMC}{Quantech simulations repository}~\cite{cp-afqmc_code}.

\bibliographystyle{quantum}
\bibliography{Bibtex}

\begin{thebibliography}{10}

\bibitem{Intro_Hubbard}
Daniel~P. Arovas, Erez Berg, Steven~A. Kivelson, and Srinivas Raghu.
\newblock ``The {H}ubbard {M}odel''.
\newblock
  \href{https://dx.doi.org/10.1146/annurev-conmatphys-031620-102024}{Annu. Rev.
  Condens. Matter Phys. {\bf 13}, 239--274}~(2022).

\bibitem{Lieb}
Elliott~H. Lieb.
\newblock ``The {H}ubbard model: Some rigorous results and open problems''.
\newblock \href{https://dx.doi.org/10.1007/978-3-662-06390-3_4}{Pages 59--77}.
\newblock Springer Berlin Heidelberg. ~(2004).

\bibitem{Bethe1}
D.C. Mattis.
\newblock ``The many-body problem: An encyclopedia of exactly solved models in
  one dimension''.
\newblock World Scientific. ~(1993).
\newblock  url:~\url{https://books.google.nl/books?id=BGdHpCAMiLgC}.

\bibitem{Bethe2}
Elliott~H. Lieb and Werner Liniger.
\newblock ``Exact analysis of an interacting bose gas. i. the general solution
  and the ground state''.
\newblock \href{https://dx.doi.org/10.1103/PhysRev.130.1605}{Phys. Rev. {\bf
  130}, 1605--1616}~(1963).

\bibitem{Chiu2018}
Christie~S. Chiu, Geoffrey Ji, Anton Mazurenko, Daniel Greif, and Markus
  Greiner.
\newblock ``Quantum {S}tate {E}ngineering of a {H}ubbard {S}ystem with
  {U}ltracold {F}ermions''.
\newblock \href{https://dx.doi.org/10.1103/PhysRevLett.120.243201}{Phys. Rev.
  Lett. {\bf 120}, 243201}~(2018).

\bibitem{Tusi2022}
D.~Tusi, L.~Franchi, L.~F. Livi, K.~Baumann, D.~Benedicto Orenes, L.~Del Re,
  R.~E. Barfknecht, T.-W. Zhou, M.~Inguscio, G.~Cappellini, M.~Capone,
  J.~Catani, and L.~Fallani.
\newblock ``Flavour-selective localization in interacting lattice fermions''.
\newblock \href{https://dx.doi.org/10.1038/s41567-022-01726-5}{Nat. Phys. {\bf
  18}, 1201--1205}~(2022).

\bibitem{Yang2022}
Jin Yang, Liyu Liu, Jirayu Mongkolkiattichai, and Peter Schauss.
\newblock ``Site-{R}esolved {I}maging of {U}ltracold {F}ermions in a
  {T}riangular-{L}attice {Q}uantum {G}as {M}icroscope''.
\newblock \href{https://dx.doi.org/10.1103/PRXQuantum.2.020344}{PRX Quantum
  {\bf 2}, 020344}~(2021).

\bibitem{Chiu2019}
Christie~S. Chiu, Geoffrey Ji, Annabelle Bohrdt, Muqing Xu, Michael Knap,
  Eugene Demler, Fabian Grusdt, Markus Greiner, and Daniel Greif.
\newblock ``String patterns in the doped {H}ubbard model''.
\newblock \href{https://dx.doi.org/10.1126/science.aav3587}{Science {\bf 365},
  251--256}~(2019).

\bibitem{spinliq}
Aaron Szasz, Johannes Motruk, Michael~P. Zaletel, and Joel~E. Moore.
\newblock ``Chiral spin liquid phase of the triangular lattice hubbard model: A
  density matrix renormalization group study''.
\newblock \href{https://dx.doi.org/10.1103/PhysRevX.10.021042}{Phys. Rev. X
  {\bf 10}, 021042}~(2020).

\bibitem{dmrg2}
Tomonori Shirakawa, Takami Tohyama, Jure Kokalj, Sigetoshi Sota, and Seiji
  Yunoki.
\newblock ``Ground-state phase diagram of the triangular lattice {H}ubbard
  model by the density-matrix renormalization group method''.
\newblock \href{https://dx.doi.org/10.1103/PhysRevB.96.205130}{Phys. Rev. B
  {\bf 96}, 205130}~(2017).

\bibitem{sza21}
Aaron Szasz and Johannes Motruk.
\newblock ``Phase diagram of the anisotropic triangular lattice {H}ubbard
  model''.
\newblock \href{https://dx.doi.org/10.1103/PhysRevB.103.235132}{Phys. Rev. B
  {\bf 103}, 235132}~(2021).

\bibitem{Zhu2022}
Zheng Zhu, D.~N. Sheng, and Ashvin Vishwanath.
\newblock ``Doped {M}ott insulators in the triangular-lattice {H}ubbard
  model''.
\newblock \href{https://dx.doi.org/10.1103/PhysRevB.105.205110}{Phys. Rev. B
  {\bf 105}, 205110}~(2022).

\bibitem{gar22}
Davis Garwood, Jirayu Mongkolkiattichai, Liyu Liu, Jin Yang, and Peter Schauss.
\newblock ``Site-resolved observables in the doped spin-imbalanced triangular
  {H}ubbard model''.
\newblock \href{https://dx.doi.org/10.1103/PhysRevA.106.013310}{Phys. Rev. A
  {\bf 106}, 013310}~(2022).

\bibitem{Becca}
Vito Marino, Federico Becca, and Luca~F. Tocchio.
\newblock ``{Stripes in the extended $t-t^\prime$ Hubbard model: A Variational
  Monte Carlo analysis}''.
\newblock \href{https://dx.doi.org/10.21468/SciPostPhys.12.6.180}{SciPost Phys.
  {\bf 12}, 180}~(2022).

\bibitem{Becca2}
Luca~F. Tocchio, Arianna Montorsi, and Federico Becca.
\newblock ``Magnetic and spin-liquid phases in the frustrated
  $t\ensuremath{-}{t}^{\ensuremath{'}}$ hubbard model on the triangular
  lattice''.
\newblock \href{https://dx.doi.org/10.1103/PhysRevB.102.115150}{Phys. Rev. B
  {\bf 102}, 115150}~(2020).

\bibitem{toc21}
Luca~F. Tocchio, Arianna Montorsi, and Federico Becca.
\newblock ``Hubbard model on triangular $n$-leg cylinders: {C}hiral and
  nonchiral spin liquids''.
\newblock \href{https://dx.doi.org/10.1103/PhysRevResearch.3.043082}{Phys. Rev.
  Research {\bf 3}, 043082}~(2021).

\bibitem{qmc1}
Yukitoshi Motome and Masatoshi Imada.
\newblock ``A {Q}uantum {M}onte {C}arlo method and its applications to
  multi-orbital {H}ubbard models''.
\newblock \href{https://dx.doi.org/10.1143/jpsj.66.1872}{J. Phys. Soc. Jpn.
  {\bf 66}, 1872--1875}~(1997).

\bibitem{Vinicius}
Vinicius Zampronio and Tommaso Macr{\`{\i} }.
\newblock ``Chiral superconductivity in the doped triangular-lattice
  {F}ermi-{H}ubbard model in two dimensions''.
\newblock \href{https://dx.doi.org/10.22331/q-2023-07-20-1061}{Quantum {\bf 7},
  1061}~(2023).

\bibitem{wie21}
Alexander Wietek, Riccardo Rossi, Fedor \ifmmode~\check{S}\else
  \v{S}\fi{}imkovic, Marcel Klett, Philipp Hansmann, Michel Ferrero, E.~Miles
  Stoudenmire, Thomas Sch\"afer, and Antoine Georges.
\newblock ``Mott {I}nsulating {S}tates with {C}ompeting {O}rders in the
  {T}riangular {L}attice {H}ubbard {M}odel''.
\newblock \href{https://dx.doi.org/10.1103/PhysRevX.11.041013}{Phys. Rev. X
  {\bf 11}, 041013}~(2021).

\bibitem{waves}
Jie Xu, Chia-Chen Chang, Eric~J Walter, and Shiwei Zhang.
\newblock ``Spin-and charge-density waves in the {H}artree{\textendash}{F}ock
  ground state of the two-dimensional {H}ubbard model''.
\newblock \href{https://dx.doi.org/10.1088/0953-8984/23/50/505601}{J. Phys.:
  Condens. Matter {\bf 23}, 505601}~(2011).

\bibitem{shi17}
Tomonori Shirakawa, Takami Tohyama, Jure Kokalj, Sigetoshi Sota, and Seiji
  Yunoki.
\newblock ``Ground-state phase diagram of the triangular lattice {H}ubbard
  model by the density-matrix renormalization group method''.
\newblock \href{https://dx.doi.org/10.1103/PhysRevB.96.205130}{Phys. Rev. B
  {\bf 96}, 205130}~(2017).

\bibitem{che22}
Bin-Bin Chen, Ziyu Chen, Shou-Shu Gong, D.~N. Sheng, Wei Li, and Andreas
  Weichselbaum.
\newblock ``Quantum spin liquid with emergent chiral order in the
  triangular-lattice {H}ubbard model''.
\newblock \href{https://dx.doi.org/10.1103/PhysRevB.106.094420}{Phys. Rev. B
  {\bf 106}, 094420}~(2022).

\bibitem{hightsuper2}
Elbio Dagotto.
\newblock ``Correlated electrons in high-temperature superconductors''.
\newblock \href{https://dx.doi.org/10.1103/RevModPhys.66.763}{Rev. Mod. Phys.
  {\bf 66}, 763--840}~(1994).

\bibitem{Kempkes}
S.~N. Kempkes, M.~R. Slot, S.~E. Freeney, S.~J.~M. Zevenhuizen,
  D.~Vanmaekelbergh, I.~Swart, and C.~Morais Smith.
\newblock ``Design and characterization of electrons in a fractal geometry''.
\newblock \href{https://dx.doi.org/10.1038/s41567-018-0328-0}{Nat. Phys. {\bf
  15}, 127--131}~(2019).

\bibitem{quantum_transport}
Xiao-Yun Xu, Xiao-Wei Wang, Dan-Yang Chen, Cristiane Morais~Smith, and Xian-Min
  Jin.
\newblock ``Quantum transport in fractal networks''.
\newblock \href{https://dx.doi.org/10.1038/s41566-021-00845-4}{Nat. Phot. {\bf
  15}, 1--8}~(2021).

\bibitem{Floq_theor1}
Shriya Pai and Abhinav Prem.
\newblock ``Topological states on fractal lattices''.
\newblock \href{https://dx.doi.org/10.1103/PhysRevB.100.155135}{Phys. Rev. B
  {\bf 100}, 155135}~(2019).

\bibitem{Floq_theor2}
Sourav Manna, Snehasish Nandy, and Bitan Roy.
\newblock ``Higher-order topological phases on fractal lattices''.
\newblock \href{https://dx.doi.org/10.1103/PhysRevB.105.L201301}{Phys. Rev. B
  {\bf 105}, L201301}~(2022).

\bibitem{Floq_exp}
Junkai Li, Qingyang Mo, Jian-Hua Jiang, and Zhaoju Yang.
\newblock ``Higher-order topological phase in an acoustic fractal
  lattice''~(2022).
\newblock  \href{http://arxiv.org/abs/2205.05298}{arXiv:2205.05298}.

\bibitem{Robert}
R.~Canyellas, Chen Liu, R.~Arouca, L.~Eek, Guanyong Wang, Yin Yin, Dandan Guan,
  Yaoyi Li, Shiyong Wang, Hao Zheng, Canhua Liu, Jinfeng Jia, and C.~Morais
  Smith.
\newblock ``Topological edge and corner states in bi fractals on insb''~(2023).
\newblock  \href{http://arxiv.org/abs/2309.09860}{arXiv:2309.09860}.

\bibitem{sign}
E.~Y. Loh, J.~E. Gubernatis, R.~T. Scalettar, S.~R. White, D.~J. Scalapino, and
  R.~L. Sugar.
\newblock ``Sign problem in the numerical simulation of many-electron
  systems''.
\newblock \href{https://dx.doi.org/10.1103/PhysRevB.41.9301}{Phys. Rev. B {\bf
  41}, 9301--9307}~(1990).

\bibitem{sign2}
Matthias Troyer and Uwe-Jens Wiese.
\newblock ``Computational complexity and fundamental limitations to fermionic
  quantum {M}onte {C}arlo simulations''.
\newblock \href{https://dx.doi.org/10.1103/PhysRevLett.94.170201}{Phys. Rev.
  Lett. {\bf 94}, 170201}~(2005).

\bibitem{Leykam2018AP31473052ArtificialFlatBandSystems}
Daniel Leykam, Alexei Andreanov, and Sergej Flach.
\newblock ``Artificial flat band systems: From lattice models to experiments''.
\newblock \href{https://dx.doi.org/10.1080/23746149.2018.1473052}{Adv. Phys.
  {\bf 3}, 1473052}~(2018).

\bibitem{Nguyen}
Huy Nguyen, Hao Shi, Jie Xu, and Shiwei Zhang.
\newblock ``{CPMC}-lab: A matlab package for constrained path {M}onte {C}arlo
  calculations''.
\newblock \href{https://dx.doi.org/10.1016/j.cpc.2014.08.003}{Comp. Phys.
  Commu. {\bf 185}, 3344--3357}~(2014).

\bibitem{Rontgen2019PRL123080504QuantumNetworkTransferStorage}
Malte R{\"o}ntgen, C.~V. Morfonios, I.~Brouzos, F.~K. Diakonos, and
  P.~Schmelcher.
\newblock ``Quantum network transfer and storage with compact localized states
  induced by local symmetries''.
\newblock \href{https://dx.doi.org/10.1103/PhysRevLett.123.080504}{Phys. Rev.
  Lett. {\bf 123}, 080504}~(2019).

\bibitem{Kempkes2023QF21CompactLocalizedBoundaryStates}
S.~N. Kempkes, P.~Capiod, S.~Ismaili, J.~Mulkens, L.~Eek, I.~Swart, and
  C.~Morais~Smith.
\newblock ``Compact localized boundary states in a quasi-{{1D}} electronic
  diamond-necklace chain''.
\newblock \href{https://dx.doi.org/10.1007/s44214-023-00026-0}{Quantum Front
  {\bf 2}, 1}~(2023).

\bibitem{Leykam2018AP3070901PerspectivePhotonicFlatbands}
Daniel Leykam and Sergej Flach.
\newblock ``Perspective: Photonic flatbands''.
\newblock \href{https://dx.doi.org/10.1063/1.5034365}{APL Phot. {\bf 3},
  070901}~(2018).

\bibitem{Sutherland1986PRB345208LocalizationElectronicWaveFunctions}
Bill Sutherland.
\newblock ``Localization of electronic wave functions due to local topology''.
\newblock \href{https://dx.doi.org/10.1103/PhysRevB.34.5208}{Phys. Rev. B {\bf
  34}, 5208--5211}~(1986).

\bibitem{Lieb1989PRL621201TwoTheoremsHubbardModel}
Elliott~H. Lieb.
\newblock ``Two theorems on the {{Hubbard}} model''.
\newblock \href{https://dx.doi.org/10.1103/PhysRevLett.62.1201}{Phys. Rev.
  Lett. {\bf 62}, 1201--1204}~(1989).

\bibitem{Cotton1990ChemicalApplicationsGroupTheory}
F.~Albert Cotton.
\newblock ``Chemical applications of group theory''.
\newblock \href{https://dx.doi.org/10.1021/ed041p113.2}{{Wiley-Interscience}}.
  {New York}~(1990).
\newblock 3 edition.

\bibitem{Landau}
L.~D. Landau and L.~M. Lifshitz.
\newblock ``Quantum mechanics non-relativistic theory, third edition: Volume
  3''.
\newblock Butterworth-Heinemann. ~(1981).
\newblock 3 edition.
\newblock  url:~\url{http://www.worldcat.org/isbn/0750635398}.

\bibitem{Kane2005PRL95146802Z2TopologicalOrder}
C.~L. Kane and E.~J. Mele.
\newblock ``Z2 topological order and the quantum spin {H}all effect''.
\newblock \href{https://dx.doi.org/10.1103/PhysRevLett.95.146802}{Phys. Rev.
  Lett. {\bf 95}, 146802}~(2005).

\bibitem{vanGelderen2010PRB81125435RashbaIntrinsicSpinorbitInteractions}
Ralph {van Gelderen} and C.~Morais Smith.
\newblock ``Rashba and intrinsic spin-orbit interactions in biased bilayer
  graphene''.
\newblock \href{https://dx.doi.org/10.1103/PhysRevB.81.125435}{Phys. Rev. B
  {\bf 81}, 125435}~(2010).

\bibitem{EDnew}
Alexander Weiße and H.~Fehske.
\newblock ``Exact diagonalization techniques''.
\newblock In {C}omputational {M}any {B}ody {P}hysics.
\newblock \href{https://dx.doi.org/10.1007/978-3-540-74686-7_18}{Pages
  529--544}.
\newblock Springer International Publishing, {Berlin/Heidelberg}~(2008).

\bibitem{Fazekas}
Patrick Fazekas.
\newblock ``Lecture notes on electron correlation and magnetism''.
\newblock \href{https://dx.doi.org/10.1142/2945}{World Scientific, 1999}.
  ~(1999).

\bibitem{Zhang}
Shiwei Zhang.
\newblock ``Ab initio electronic structure calculations by auxiliary-field
  quantum {M}onte {C}arlo''.
\newblock In Handbook of Materials Modeling.
\newblock \href{https://dx.doi.org/10.1007/978-3-319-42913-7_47-1}{Pages
  1--27}.
\newblock Springer International Publishing~(2018).

\bibitem{Zhang_bosons}
Wirawan Purwanto and Shiwei Zhang.
\newblock ``Quantum monte carlo method for the ground state of many-boson
  systems''.
\newblock \href{https://dx.doi.org/10.1103/PhysRevE.70.056702}{Phys. Rev. E
  {\bf 70}, 056702}~(2004).

\bibitem{Zhang_symm}
Hao Shi and Shiwei Zhang.
\newblock ``Symmetry in auxiliary-field quantum monte carlo calculations''.
\newblock \href{https://dx.doi.org/10.1103/PhysRevB.88.125132}{Phys. Rev. B
  {\bf 88}, 125132}~(2013).

\bibitem{Pedersen2020}
Thomas~Garm Pedersen.
\newblock ``{G}raphene fractals: {E}nergy gap and spin polarization''.
\newblock \href{https://dx.doi.org/10.1103/PhysRevB.101.235427}{Phys. Rev. B
  {\bf 101}, 235427}~(2020).

\bibitem{honeycomb_useful}
Marcin Raczkowski, Robert Peters, Th\d{i}~Thu Ph\`ung, Nayuta Takemori,
  Fakher~F. Assaad, Andreas Honecker, and Javad Vahedi.
\newblock ``Hubbard model on the honeycomb lattice: From static and dynamical
  mean-field theories to lattice quantum monte carlo simulations''.
\newblock \href{https://dx.doi.org/10.1103/PhysRevB.101.125103}{Phys. Rev. B
  {\bf 101}, 125103}~(2020).

\bibitem{Birkam_Rice}
W.~F. Brinkman and T.~M. Rice.
\newblock ``Application of gutzwiller's variational method to the
  metal-insulator transition''.
\newblock \href{https://dx.doi.org/10.1103/PhysRevB.2.4302}{Phys. Rev. B {\bf
  2}, 4302--4304}~(1970).

\bibitem{Sorella}
S.~Sorella and E.~Tosatti.
\newblock ``Semi-metal-insulator transition of the hubbard model in the
  honeycomb lattice''.
\newblock \href{https://dx.doi.org/10.1209/0295-5075/19/8/007}{EPL {\bf 19},
  699}~(1992).

\bibitem{Raczkowski2020}
Marcin Raczkowski, Robert Peters, Th\d{i}~Thu Ph\`ung, Nayuta Takemori,
  Fakher~F. Assaad, Andreas Honecker, and Javad Vahedi.
\newblock ``{H}ubbard model on the honeycomb lattice: {F}rom static and
  dynamical mean-field theories to lattice quantum {M}onte {C}arlo
  simulations''.
\newblock \href{https://dx.doi.org/10.1103/PhysRevB.101.125103}{Phys. Rev. B
  {\bf 101}, 125103}~(2020).

\bibitem{Pal2018PRB97195101FlatBandsFractallikeGeometry}
Biplab Pal and Kush Saha.
\newblock ``Flat bands in fractal-like geometry''.
\newblock \href{https://dx.doi.org/10.1103/PhysRevB.97.195101}{Phys. Rev. B
  {\bf 97}, 195101}~(2018).

\bibitem{Biswas2023PELSaN153115762DesignerQuantumStatesFractal}
Sougata Biswas and Arunava Chakrabarti.
\newblock ``Designer quantum states on a fractal substrate: {{Compact}}
  localization, flat bands and the edge modes''.
\newblock \href{https://dx.doi.org/10.1016/j.physe.2023.115762}{Physica E {\bf
  153}, 115762}~(2023).

\bibitem{Nandy2021PS96045802ControlledImprisonmentWavePacket}
Atanu Nandy.
\newblock ``Controlled imprisonment of wave packet and flat bands in a fractal
  geometry''.
\newblock \href{https://dx.doi.org/10.1088/1402-4896/abdcf6}{Phys. Scr. {\bf
  96}, 045802}~(2021).

\bibitem{Xie2021AP6116104FractallikePhotonicLatticesLocalized}
Yuqing Xie, Limin Song, Wenchao Yan, Shiqi Xia, Liqin Tang, Daohong Song,
  Jun-Won Rhim, and Zhigang Chen.
\newblock ``Fractal-like photonic lattices and localized states arising from
  singular and nonsingular flatbands''.
\newblock \href{https://dx.doi.org/10.1063/5.0068032}{APL Phot. {\bf 6},
  116104}~(2021).

\bibitem{Hanafi2022AOM102102523LocalizedStatesEmergingSingular}
Haissam Hanafi, Philip Menz, and Cornelia Denz.
\newblock ``Localized {{States Emerging}} from {{Singular}} and {{Nonsingular
  Flat Bands}} in a {{Frustrated Fractal-Like Photonic Lattice}}''.
\newblock \href{https://dx.doi.org/10.1002/adom.202102523}{Adv. Opt. Mater.
  {\bf 10}, 2102523}~(2022).

\bibitem{Westerhout}
Tom Westerhout, Edo van Veen, Mikhail~I. Katsnelson, and Shengjun Yuan.
\newblock ``Plasmon confinement in fractal quantum systems''.
\newblock \href{https://dx.doi.org/10.1103/PhysRevB.97.205434}{Phys. Rev. B
  {\bf 97}, 205434}~(2018).

\bibitem{Iliasov}
Askar~A. Iliasov, Mikhail~I. Katsnelson, and Andrey~A. Bagrov.
\newblock ``Strong enhancement of superconductivity on finitely ramified
  fractal lattices''~(2024).
\newblock  \href{http://arxiv.org/abs/2310.11497}{arXiv:2310.11497}.

\bibitem{Manna}
Sourav Manna, Biplab Pal, Wei Wang, and Anne E.~B. Nielsen.
\newblock ``Anyons and fractional quantum hall effect in fractal dimensions''.
\newblock \href{https://dx.doi.org/10.1103/PhysRevResearch.2.023401}{Phys. Rev.
  Res. {\bf 2}, 023401}~(2020).

\bibitem{Jaworowski}
Blazej Jaworowski, Michael Iversen, and Anne E.~B. Nielsen.
\newblock ``Approximate {H}ofstadter- and {K}apit-{M}ueller-like parent
  {H}amiltonians for {L}aughlin states on fractals''.
\newblock \href{https://dx.doi.org/10.1103/PhysRevA.107.063315}{Phys. Rev. A
  {\bf 107}, 063315}~(2023).

\bibitem{Ferri}
Akihisa Koga and Sam Coates.
\newblock ``Ferrimagnetically ordered states in the {H}ubbard model on the
  hexagonal golden-mean tiling''.
\newblock \href{https://dx.doi.org/10.1103/PhysRevB.105.104410}{Phys. Rev. B
  {\bf 105}, 104410}~(2022).

\bibitem{Lieb1968}
Elliott~H. Lieb and F.~Y. Wu.
\newblock ``{A}bsence of {M}ott {T}ransition in an {E}xact {S}olution of the
  {S}hort-{R}ange, {O}ne-{B}and {M}odel in {O}ne {D}imension''.
\newblock \href{https://dx.doi.org/10.1103/PhysRevLett.20.1445}{Phys. Rev.
  Lett. {\bf 20}, 1445--1448}~(1968).

\bibitem{Haldane1980}
F.~D.~M. Haldane.
\newblock ``{G}eneral {R}elation of {C}orrelation {E}xponents and {S}pectral
  {P}roperties of {O}ne-{D}imensional {F}ermi {S}ystems: {A}pplication to the
  {A}nisotropic $s=\frac{1}{2}$ {H}eisenberg {C}hain''.
\newblock \href{https://dx.doi.org/10.1103/PhysRevLett.45.1358}{Phys. Rev.
  Lett. {\bf 45}, 1358--1362}~(1980).

\bibitem{Haldane1981}
F~D~M Haldane.
\newblock ``{\textquotesingle}{L}uttinger liquid theory{\textquotesingle} of
  one-dimensional quantum fluids. {I}. {P}roperties of the {L}uttinger model
  and their extension to the general 1{D} interacting spinless {F}ermi gas''.
\newblock \href{https://dx.doi.org/10.1088/0022-3719/14/19/010}{J. Phys. C:
  Solid State Phys. {\bf 14}, 2585--2609}~(1981).

\bibitem{Haldane1981-2}
F.D.M. Haldane.
\newblock ``{D}emonstration of the {\textquotedblleft}{L}uttinger
  liquid{\textquotedblright} character of {B}ethe-ansatz-soluble models of
  1-{D} quantum fluids''.
\newblock \href{https://dx.doi.org/10.1016/0375-9601(81)90049-9}{Phys. Lett. A
  {\bf 81}, 153--155}~(1981).

\bibitem{cp-afqmc_code}
V.~Zampronio.
\newblock ``Cp-afqmc''~(2022).

\bibitem{zha95}
Shiwei Zhang, J.~Carlson, and J.~E. Gubernatis.
\newblock ``Constrained {P}ath {Q}uantum {M}onte {C}arlo {M}ethod for {F}ermion
  {G}round {S}tates''.
\newblock \href{https://dx.doi.org/10.1103/PhysRevLett.74.3652}{Phys. Rev.
  Lett. {\bf 74}, 3652--3655}~(1995).

\bibitem{zha97}
Shiwei Zhang, J.~Carlson, and J.~E. Gubernatis.
\newblock ``Constrained path {M}onte {C}arlo method for fermion ground
  states''.
\newblock \href{https://dx.doi.org/10.1103/PhysRevB.55.7464}{Phys. Rev. B {\bf
  55}, 7464--7477}~(1997).

\bibitem{tro59}
H.~F. Trotter.
\newblock ``On the product of semi-groups of operators''.
\newblock
  \href{https://dx.doi.org/https://doi.org/10.1090/S0002-9939-1959-0108732-6}{Proc.
  Amer. Math. Soc. {\bf 10}, 545--551}~(1959).

\bibitem{hir85}
J.~E. Hirsch.
\newblock ``Two-dimensional {H}ubbard model: {N}umerical simulation study''.
\newblock \href{https://dx.doi.org/10.1103/PhysRevB.31.4403}{Phys. Rev. B {\bf
  31}, 4403--4419}~(1985).

\bibitem{rey82}
Peter~J. Reynolds, David~M. Ceperley, Berni~J. Alder, and William~A. Lester.
\newblock ``Fixed-node quantum {M}onte {C}arlo for molecules a)-b)''.
\newblock \href{https://dx.doi.org/10.1063/1.443766}{J. Chem. Phys. {\bf 77},
  5593--5603}~(1982).
\newblock
  \href{http://arxiv.org/abs/https://doi.org/10.1063/1.443766}{arXiv:https://doi.org/10.1063/1.443766}.

\bibitem{qin16}
Mingpu Qin, Hao Shi, and Shiwei Zhang.
\newblock ``Benchmark study of the two-dimensional {H}ubbard model with
  auxiliary-field quantum {M}onte {C}arlo method''.
\newblock \href{https://dx.doi.org/10.1103/PhysRevB.94.085103}{Phys. Rev. B
  {\bf 94}, 085103}~(2016).

\bibitem{zha91}
X.~Y. Zhang, Elihu Abrahams, and G.~Kotliar.
\newblock ``{Q}uantum {M}onte {C}arlo algorithm for constrained fermions:
  {A}pplication to the infinite-${U}$ {H}ubbard model''.
\newblock \href{https://dx.doi.org/10.1103/PhysRevLett.66.1236}{Phys. Rev.
  Lett. {\bf 66}, 1236--1239}~(1991).

\bibitem{pur04}
Wirawan Purwanto and Shiwei Zhang.
\newblock ``Quantum {M}onte {C}arlo method for the ground state of many-boson
  systems''.
\newblock \href{https://dx.doi.org/10.1103/PhysRevE.70.056702}{Phys. Rev. E
  {\bf 70}, 056702}~(2004).

\end{thebibliography}

\onecolumn
\appendix

\section{CP-AFQMC method}\label{app:cpmc}

Consider an initial state \( |\Psi(0)\rangle \) that is not orthogonal to the ground state \( |\Psi_0\rangle \) of the Hamiltonian \( H \). Through imaginary-time evolution, \( |\Psi(\tau)\rangle = \exp(-\tau H)|\Psi(0)\rangle \) asymptotically converges to \( |\Psi_0\rangle \) as \( \tau \) (a real number) increases. In the Auxiliary Field Quantum Monte Carlo (AFQMC) method, the antisymmetric wave function is expressed as a linear combination of Slater determinants,
\begin{equation}\label{eq_basis}
\left|\Psi(\tau)\right> = \sum_k \xi(\Phi_k)\left|\Phi_k(\tau)\right>.
\end{equation}
In our simulations, the coefficients \( \xi(\Phi_k) \) are not explicitly considered. As imaginary-time evolution proceeds, Slater determinants are either replicated or eliminated. The number of a given \( |\Phi_k\rangle \) in the sum reflects \( \xi(\Phi_k) \)~\cite{zha95,zha97}.

Typically, the process starts at \( \tau = 0 \) with all Slater determinants equal to a given trial state \( |\Phi_T\rangle \), which approximates the ground state and is usually derived from mean-field theories. These determinants are then updated via the application of the imaginary-time evolution operator in a stochastic process. For large systems, direct diagonalization of the Fermi-Hubbard Hamiltonian becomes impractical. Consequently, the Trotter formula~\cite{tro59} is employed to factorize the evolution operator into a product of three terms
\begin{equation}\label{eq_tf}
{\rm e}^{-\delta\tau H}={\rm e}^{-\frac{\delta\tau}{2}K}{\rm e}^{-\delta\tau V}{\rm e}^{-\frac{\delta\tau}{2}K} + {\cal O}(\delta\tau^2),
\end{equation}
where \( K \) and \( V \) represent the hopping and interaction terms, respectively, in the Fermi-Hubbard Hamiltonian. By choosing a sufficiently small \( \delta\tau \), the error introduced by neglecting the \( {\cal O}(\delta\tau^2) \) terms in Eq.~\eqref{eq_tf} can be minimized to be less than the statistical uncertainty inherent to Monte Carlo calculations, thereby maintaining numerical exactness.

The desired limit \( \tau = n\delta\tau \gg t^{-1} \), where \( t \) denotes the hopping strength, is achieved after \( n \) successive applications of this small-\( \delta\tau \) approximation to \( |\Psi(0)\rangle \). A specific iteration on the Slater determinants is described by
\begin{equation}\label{eq:it}
|\Phi^{n+1}_k\rangle = {\rm e}^{-\frac{\delta\tau}{2}K}{\rm e}^{-\delta\tau V}{\rm e}^{-\frac{\delta\tau}{2}K} |\Phi^{n}_k\rangle,
\end{equation}
where the superscript \( n \) indicates the imaginary time \( \tau = n\delta\tau \).

The application of one-body operators on \( |\Phi_k^n\rangle \) results in another Slater determinant. Consequently, \( \exp(-\delta\tau K/2) \) propagates \( |\Phi_k^n\rangle \). However, since \( V \) is a sum of two-body operators, the remaining exponential presents a challenge. To address this, the Hubbard-Stratonovich transformation is employed to convert the two-body interaction into one-body interactions between electrons and auxiliary fields \( x \). We adopt the spin discrete decomposition~\cite{hir85}
\begin{equation}\label{eq_dhsd}
{\rm e}^{-\delta\tau n_{i\uparrow}n_{i\downarrow}}=
{\rm e}^{-\frac{\delta\tau}{2}U(n_{i\uparrow}+n_{i\downarrow})}
\sum_{x=\pm 1} p(x)
{\rm e}^{\gamma x (n_{i\uparrow}-n_{i\downarrow})},
\end{equation}
where \( p(x)=1/2 \) and \( \gamma \) is defined by the relation \( \cosh(\gamma)=\exp(\delta\tau U/2) \).

For the Fermi-Hubbard Hamiltonian \( H \), this is expressed as
\begin{equation}\label{eq_dm}
{\rm e}^{\delta\tau H} \approx
\sum_{\bm{x}} p(\bm{x})
{\rm e}^{-\frac{\beta}{2}K}
B_V(\bm{x})
{\rm e}^{-\frac{\beta}{2}K},
\end{equation}
where \( \bm{x}=(x_1,x_2,\ldots,x_M) \) represents a configuration of auxiliary fields, with \( M \) being the number of lattice sites, to be sampled within Monte Carlo calculations. Here, \( p(\bm{x})=(1/2)^M \) is a probability distribution function (pdf), and \( B_V(\bm{x}) \) is a product of one-body exponentials
\begin{equation}\label{eq_b}
B_V(\bm{x})=
\prod_i {\rm e}^{-\frac{\delta\tau}{2}U(n_{i\uparrow}+n_{i\downarrow})+\gamma x_i (n_{i\uparrow}-n_{i\downarrow})}.
\end{equation}

Direct simulation of Eq.~\eqref{eq:it} is inefficient due to the constant nature of \( p(\bm{x}) \), necessitating an importance sampling technique~\cite{zha95,zha97}. Importance sampling also aids in defining an estimator for system properties and determining constraints that eliminate the sign problem. The importance function implemented is \( O_T(\Phi_k^n)=\langle \Phi_T|\Phi_k^n\rangle \), leading to the modified imaginary-time evolution
\begin{equation}\label{eq:it_is}
|\tilde{\Psi}^{n+1}\rangle = \sum_{\bm{x}}\tilde{p}(\bm{x}){\rm e}^{-\frac{\delta\tau}{2}K}B_V(\bm{x}){\rm e}^{-\frac{\delta\tau}{2}K}|\tilde{\Psi}^n\rangle,
\end{equation}
with the modified pdf \( \tilde{p}(\bm{x})=O_T(\Phi_j^n)p(\bm{x})/O_T(\Phi_j^{n-1}) \).

Since the pdf \( \tilde{p}(\bm{x}) \) is usually not normalized, the normalization factor for each Slater determinant \( N(\Phi_k^n) \) is defined, transforming the iterative projection equation into
\begin{equation}\label{eq_is}
\left|\Phi^n_k\right>=N(\Phi_k^n)\sum_{\bm{x}}
\frac{\tilde{p}(\bm{x})}{N(\Phi_k^n)}{\rm e}^{-\frac{\delta\tau}{2}K}B_V {\rm e}^{-\frac{\delta\tau}{2}K}
\left|\Phi^{n-1}_k\right>.
\end{equation}

To manage normalization, weights for each Slater determinant \( |\tilde{\Psi^n}\rangle = \sum_k \omega_k^n|\Phi_k^n\rangle \) are introduced, updating the weights in each iteration as \( \omega^n_k = N(\Phi_k^n)\omega^{n-1}_k \), with \( \omega_k^0=1 \). In practice, the pdf \( \tilde{p}(\bm{x}) \) is sampled by considering each auxiliary field in the configuration \( \bm{x} \) individually. Detailed implementation of the sampling and weight update process can be found in Ref.~\cite{Nguyen}.

The equivalence \( \left|\Phi_k^n\right>=-\left|\Phi_k^n\right> \) results in a sign problem that impedes numerical convergence. To mitigate this, auxiliary-field paths are constrained to a region of the configuration space where \( O_T(\Phi_k^n) > 0 \), similar to the fixed-node approximation~\cite{rey82}. Since the method remains numerically exact if the nodal structure of the trial wave function matches the ground state, this constraint is effective. However, the exact nodal structure of the ground state is generally unknown, necessitating the use of \( \left|\Phi_T\right> \) as an approximation, introducing a small systematic error~\cite{zha95,zha97,Nguyen,qin16}.

Ground-state estimates of the total energy are obtained using the mixed estimator
\begin{equation}\label{eq_mixed}
\langle H \rangle_{\text{mix}} =
\frac{\sum_k \omega_k^n E_k^n}{\sum_j \omega_k^n},
\end{equation}
with $E^n_k=\langle\Phi_T|H|\Phi^n_k\rangle / O_T(\Phi_k^n)$ and sufficiently large \( n \). This mixed estimator is exact only if the operator in the numerator of Eq.~\eqref{eq_mixed} commutes with the Hamiltonian \( H \).

Estimates of other physical observables require the back-propagation technique~\cite{zha95,zha91,pur04}. The back-propagation estimator is derived from
\begin{equation}\label{eq_bp}
\left< O \right>_{\text{bp}} \propto \langle \Phi_T | {\rm e}^{-\tau_{\text{bp}} H} O {\rm e}^{-(\tau - \tau_{\text{bp}}) H} |\Psi(0) \rangle,
\end{equation}
which asymptotically reaches the average value of the observable \( O \) in the ground state for $\tau - \tau_{\text{bp}}$ and $\tau_{\text{bp}} \gg t^{-1}$. The numerical evaluation of Eq.~\eqref{eq_bp} is efficiently performed by storing the auxiliary fields sampled during the forward propagation $\exp(-\tau_{\text{bp}} H) |\Psi(0) \rangle$ and using them for back propagation $\langle \Phi_T |$. For a detailed description of this estimator, see Ref.~\cite{pur04}.

\section{Explanation of the six-fold symmetry of the structure factor}

\begin{figure}[h]
	\centering
	\includegraphics[width=0.3\columnwidth]{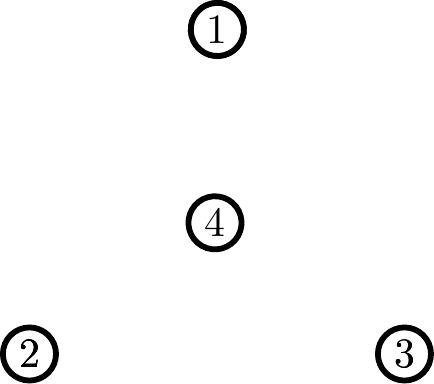}
 	\caption{An arrangement of four sites $1$, $2$, $3$, and $4$ that is invariant by a rotations of $2\pi/3$ around the central site $4$.}
	\label{fig:structureFactorExplanation}
\end{figure}

To understand the six-fold rotational symmetry of the magnetic structure factor \begin{equation*}
    S(\bm{k}) = \frac{1}{N^2} \sum_{i,j} e^{i \bm{k} \cdot \bm{r}_{ij}} c_{m}(i,j)\,,
\end{equation*} as visible in \cref{fig:struct} (a), let us start by noting that our Hamiltonian enjoys a three-fold rotational symmetry. Without loss of generality, let us restrict ourselves to an arrangement of four sites; the generalization to an arbitrary number of sites is then straightforward.
The setup is depicted in \cref{fig:structureFactorExplanation}.
To use the symmetry, we introduce the rotation operator
\begin{equation}
    R(\mathbf{x},\phi) = \begin{pmatrix} \cos(\phi) & - \sin(\phi) \\ \sin(\phi) & \cos(\phi) \end{pmatrix} \cdot \mathbf{x} := M_{\phi} \cdot \mathbf{x} \,,
\end{equation}
which rotates the vector $\mathbf{x}$ counterclockwise by an angle $\phi$ around the origin (which is here taken to be the location of site $4$ in \cref{fig:structureFactorExplanation}).
\label{app:structureFactor}

Explicitly written out, the structure factor of this arrangement reads
\begin{equation}
    S(\bm{k}) =  2c_m(1,2) \left(A_{1,2} + A_{2,3} + A_{3,1}  \right)
    + 2c_{1,4} \left(A_{1,4} + A_{2,4} + A_{3,4}  \right) 
    + \sum_{i=1}^4 c_m(i,i)\,,
\end{equation}
with $A_{i,j} = \cos(\mathbf{q} \cdot \mathbf{r}_{i,j})$, and where we have used the fact that $c_m(i,j) = c_m(j,i)$, and additionally that the setup is three-fold rotationally symmetric.
Additionally, due to this symmetry, the following relations (with $\alpha=\frac{2\pi}{3}$) hold
\begin{equation}
 \label{eq:rotationMapping}
\mathbf{r}_{2} = R(\mathbf{r}_{1},\alpha), \ \ \mathbf{r}_{3} = R(\mathbf{r}_{2},\alpha), \ \
\mathbf{r}_{1} = R(\mathbf{r}_{3},\alpha) \ \ \text{and} \ \
\mathbf{r}_{4} = R(\mathbf{r}_{4},\alpha).
\end{equation}

We are interested in the rotational symmetries of the structure factor; in particular, we want to show that $S(\bm{k}) = S(M_{\alpha/2} \mathbf{k})$.
The crucial point now is to see that the scalar product occurring in $\cos(\mathbf{k} \cdot \mathbf{r}_{i,j})$ fulfills the property 
\begin{equation}
    (M_{\alpha/2} \, \mathbf{k}) \cdot \mathbf{r}_{i,j} = - \mathbf{k} \cdot (M_{\alpha} \, \mathbf{r}_{i,j}) \,,
\end{equation}
from which it follows that
\begin{equation*}
    \cos((M_{\alpha/2} \, \mathbf{k}) \cdot \mathbf{r}_{i,j}) = \cos(\mathbf{k} \cdot (M_{\alpha} \, \mathbf{r}_{i,j}))
\end{equation*}
Then, since $M_\alpha \mathbf{r}_{i,j} = M_\alpha \mathbf{r}_i - M_\alpha \mathbf{r}_j$, we can use \cref{eq:rotationMapping} to see that
\begin{equation}
\begin{split}
    S(M_{\alpha/2} \mathbf{k}) & = 2c_m(1,2) \left(A_{2,3} + A_{3,1} + A_{1,2}  \right) 
     + 2c_{1,4} \left(A_{2,4} + A_{3,4} + A_{1,4}  \right) 
     + \sum_{i=1}^4 c_m(i,i) \\
    & = S(\bm{k}).
\end{split}
\end{equation}
In other words, $S(\mathbf{k})$ is 6-fold rotational symmetric.

\end{document}